\documentclass[fleqn,usenatbib]{mnras}

\usepackage{newtxtext,newtxmath}

\usepackage[T1]{fontenc}
\usepackage{ae,aecompl}
\usepackage{multirow}
\usepackage{xcolor}

\usepackage{graphicx}	
\usepackage{amsmath}	
\usepackage{caption}
\usepackage{subcaption}
\usepackage{threeparttable}
\usepackage{float}
\restylefloat{figure}

\title[MS to Starburst Transition in Interm-z U/LIRGs]{Unveiling the Main Sequence to Starburst Transition Region with a Sample of Intermediate Redshift Luminous Infrared Galaxies}

\author[L. Hogan et al.]{L. Hogan$^{1}$\thanks{Contact e-mail: \href{mailto:laurence.hogan@physics.ox.ac.uk}{laurence.hogan@physics.ox.ac.uk}}\thanks{Present address: Department of Physics, University of Oxford, Keble Road, Oxford OX1 3RH, UK},
 D. Rigopoulou$^{1}$, S. Garc\'{i}a-Burillo$^{2}$, A. Alonso-Herrero$^{3}$, L. Barrufet$^{4}$,  \newauthor F. Combes$^{5}$, I. Garc\'{i}a-Bernete$^{1}$, G. E. Magdis$^{6,7,8,9}$, M. Pereira-Santaella$^{1,10}$, N. Thatte$^{1}$,
  \newauthor A. Wei{\ss}$^{11}$.
\\
\\
$^{1}$Department of Physics, University of Oxford, Keble Road, Oxford OX1 3RH, UK \\
$^{2}$ Observatorio Astron\'{o}mico Nacional (OAN-IGN)-Observatorio de Madrid, Alfonso XII, 3, 28014, Madrid, Spain \\
$^{3}$ Centro de Astrobiolog\'{i}a (CAB, CSIC-INTA), ESAC Campus, 28692 Villanueva de la Cañada, Madrid, Spain \\
$^{4}$ Geneva Observatory, University of Geneva, Ch. des Mail-lettes 51, 1290 Versoix, Switzerland \\
$^{5}$LERMA, Obs. de Paris, PSL Univ., Coll\'{e}ge de France, CNRS, Sorbonne Univ., Paris, France \\
$^{6}$ Cosmic Dawn Center (DAWN), Copenhagen, Denmark \\
$^{7}$ DTU-Space, Technical University of Denmark, Elektrovej 327, DK-2800 Kgs. Lyngby, Denmark \\
$^{8}$University of Copenhagen, Lyngbyvej 2, DK-2100 Copenhagen {\O}, Denmark \\
$^{9}$ Institute for Astronomy, Astrophysics, Space Applications and Remote Sensing, National Observatory of Athens, GR-15236 Athens, Greece \\
$^{10}$Centro de Astrobiolog\'{i}a (CSIC-INTA), Ctra. de Ajalvir, Km 4, 28850, Torrej\'on de Ardoz, Madrid, Spain \\
$^{11}$ Max-Planck-Institut f\"{u}r Radioastronomie, Auf dem H\"{u}gel 69 D-53121 Bonn, Germany
}

\date{Accepted XXX. Received YYY; in original form ZZZ}

\pubyear{2021}

\begin{document}
\label{firstpage}
\pagerange{\pageref{firstpage}--\pageref{lastpage}}
\maketitle

\begin{abstract}
We present a CO(3-2) study of four systems composed of six (ultra) luminous infrared galaxies (U/LIRGs), located at 0.28 $<$ z $<$ 0.44, that straddle the transition region between regular star forming galaxies and starbursts. These galaxies benefit from previous multi-wavelength analysis allowing in depth exploration of an understudied population of U/LIRGs at a time when the universe is experiencing a rapid decline in star formation rate density. We detect CO(3-2) emission in four targets and these galaxies fall between the loci of regular star forming galaxies and starbursts on the Kennicutt-Schmidtt relation. Compared to low luminosity LIRGs and high luminosity ULIRGs at similar redshifts, we find they all have similar molecular gas budgets with the difference in their star formation rates (SFR) driven by the star formation efficiency (SFE). This suggests that at these redshifts large molecular gas reservoirs must coincide with an increased SFE to transition a galaxy into the starburst regime. We studied the structure and kinematics and found our four detections are either interacting or have disturbed morphology which may be driving the SFE. One of the CO(3-2) non-detections has a strong continuum detection, and has been previously observed in H$\alpha$, suggesting an unusual interstellar medium for a ULIRG.
We conclude that our sample of transitioning U/LIRGs fill the gap between regular star forming galaxies and starbursts, suggest a continuous change in SFE between these two populations and the increased SFE may be driven by morphology and differing stages of interaction.
\end{abstract}




\section{Introduction}
\label{section:intro}

In recent decades multi-wavelength studies have revealed the connection between gas, dust and existing stellar mass within a galaxy, and how these processes shape the evolution of galaxies through cosmic time. Observations have shown a tight correlation between the star formation rate (SFR) and stellar mass (M$_{\star}$) of a galaxy, known as the main sequence (MS) of star forming galaxies (e.g. \citealp{noeske_star_2007}, \citealp{daddi_multiwavelength_2007}), with a galaxies position on the MS plane given by the specific SFR (sSFR = SFR / M$_{\star}$). The sSFR can be used to categorize two populations of star forming galaxies: normal star forming galaxies and starbursts. Normal star forming galaxies lie within a scatter of 0.3 dex of the MS whereas starbursts fall in the region a factor of 4 or more above the MS correlation (e.g. \citealp{whitaker_star_2012}). It is suggested that normal galaxies are forming stars in a quasi steady state, fuelled from continuous in-falling gas from the intergalactic medium (e.g. \citealp{dekel_formation_2009}); whereas starbursts seem to be undergoing a short lived period of heightened star formation driven by stochastic processes such as mergers (e.g. \citealp{tacconi_submillimeter_2008}, \citealp{daddi_different_2010}. \citealp{genzel_study_2010}) The normalisation of the MS correlation increases with redshift (e.g. \citealp{speagle_highly_2014}), in line with the increase in the cosmic star formation rate density (SFRD) between z = 0 - 2  (e.g. \citealp{madau_cosmic_2014}). 

It is now widely accepted that a strong correlation exists between the surface density of gas within a galaxy and the surface density of SFR known as the Kennicutt-Schmidtt law, $\Sigma_{\rm SFR} \propto \Sigma^{N}_{\rm gas}$, with gas in local galaxies following a tight non-linear scaling law given by index N = 1.4 $\pm$ 0.15 (\citealp{schmidt_rate_1959}, \citealp{kennicutt_star_1998}). Shallower indices are found for the molecular gas mass surface density, $\Sigma_{\rm mol}$, with the approximately linear relationship suggesting a stronger link between SFR and molecular gas within a galaxy (e.g. \citealp{bigiel_star_2008}, \citealp{bigiel_constant_2011}, \citealp{schruba_molecular_2011} \citealp{leroy_molecular_2013}, \citealp{de_los_reyes_revisiting_2019}). 
The ratio of SFR and molecular gas gives the star formation efficiency (SFE) of a galaxy ($\Sigma_{\rm SFR} / \Sigma_{\rm mol}$), which tells us how quickly a galaxy can convert its current molecular gas reservoirs into stars, but does not take into account any mass recycled back into the interstellar medium (ISM) from winds and supernova remnants (e.g. \citealp{tacconi_phibss:_2018}). MS galaxies have a lower SFE than starburst galaxies (e.g. \citealp{tacconi_evolution_2020}) suggesting different triggering mechanisms are responsible for star formation in these two groups, and it is key to understand SFE and depletion time ($\tau_{\rm dep}$ $\equiv$ 1/SFE) to explain the physics of the MS. The bi-modality of regular star forming galaxies versus starbursts in the MS parameter space can be due to higher gas fractions and/or higher SFE, with observations showing that the SFE and gas fraction increase with both distance above the MS and lookback time, although broadly remain constant along the MS in a given epoch (e.g. \citealp{magdis_evolving_2012}, \citealp{genzel_combined_2015}, \citealp{scoville_evolution_2017}, \citealp{tacconi_evolution_2020}). It also appears that the gas fraction increases much more dramatically with lookback time than the SFE (e.g. \citealp{tacconi_phibss:_2018}). Quiescent galaxies, which lie below the MS, are linked with both lower SFE and gas fractions (e.g. \citealp{piotrowska_towards_2020}). 

Luminous (10$^{11}$ $<$ L$_{\rm IR}$ $\equiv$ L$_{8 - 1000 \mu m}$ $<$ 10$^{12}$ L$_{\odot}$, LIRGs) and ultra luminous infrared (IR) galaxies (10$^{12}$ $<$ L$_{\rm IR}$ $<$ 10$^{13}$ L$_{\odot}$, ULIRGs) are amongst the most intensely star forming galaxies in the universe (\citealp{sanders_luminous_1996}) with SFRs ranging from tens up to thousands of solar masses per year (e.g. \citealp{rigopoulou_multiwavelength_1996}); making them excellent probes of the interplay between gas and SFR. Their IR emission arises from dust heated by newly formed massive stars and active galactic nuclei (AGN), with AGN being more prevalent and powerful with increasing luminosity (e.g. \citealp{tran_isocam_2001}, \citealp{veilleux_spitzer_2009}, \citealp{nardini_role_2010}, \citealp{alonso-herrero_local_2012}).

Locally (z $<$ 0.2) ULIRGs are rare and appear to be undergoing a transient starburst phase, fuelled by gas rich major mergers, whereas local LIRGs show a more diverse range of morphological types, such as isolated disks and minor mergers (e.g. \citealp{kartaltepe_multiwavelength_2010}, \citealp{bellocchi_distinguishing_2016}, \citealp{larson_morphology_2016}). Their IR spectral energy distributions (SED) are dominated by thermal emission from dust and have dust temperatures in the range 30-55K (\citealp{clements_herus:_2018} and references therein). They are very efficient at converting their molecular gas into stars (e.g. \citealp{gao_star_2004}) with depletion times on the order of 10 Myr (e.g. \citealp{pereira-santaella_are_2021}). The extent of their star forming regions is relatively compact, with sizes of 0.3 - 2 kpc (e.g. \citealp{alonso-herrero_near-infrared_2006}, \citealp{pereira-santaella_spatially_2018}).

Although matched in luminosity, observations have shown that high redshift (z $>$ 1) U/LIRGs are very different from their local counterparts. Their co-moving density increases by a factor of $\sim$ 1000 between redshifts of 1 - 2, where they are responsible for as much as 50\% of the SFRD (e.g. \citealp{magnelli_evolution_2011}, \citealp{murphy_accounting_2011}). Many authors have noted differences in their SEDs (e.g. \citealp{farrah_nature_2008} \citealp{muzzin_well-sampled_2010}), physical properties and morphologies (e.g. \citealp{kartaltepe_goods-herschel_2012}, \citealp{hogan_integral_2021}). High-resolution studies have revealed that the star-forming regions of high-z U/LIRGs are relatively extended with sizes on the of order 3 - 16 kpc (e.g. \citealp{iono_luminous_2009}, \citealp{tacconi_phibss:_2013}).  Dynamically, high redshift U/LIRGs appear to be a mixture of mergers and disk galaxies, with spatially resolved observations revealing large rotating disks at z $\sim$ 2 with SFRs in the hundreds of M$_{\odot}$ yr$^{-1}$ without any sign of ongoing major merging (e.g. \citealp{forster_schreiber_sins_2009}, \citealp{wisnioski_kmos_2015}, \citealp{hogan_integral_2021}). In general, they do not exhibit the metal deficiency seen in local U/LIRGs (e.g. \citealp{gracia-carpio_far-infrared_2011}, \citealp{diaz-santos_herschel/pacs_2017}), which suggests softer radiation fields and more extended regions of star formation (e.g. \citealp{diaz-santos_herschel/pacs_2017}, \citealp{herrera-camus_shining_2018}). 

At intermediate redshifts (0.2 $<$ z $<$ 1) U/LIRGs have properties that resemble both the high-z and local U/LIRGs. Intermediate-z U/LIRGs appear to be a mixture of interacting and isolated objects, can have a lower dust temperatures than local U/LIRGs, but have a similar level of dust obscuration (\citealp{pereira-santaella_optical_2019}). They can have an SFE, dust and interstellar medium (ISM) that exhibits characteristics that more closely resemble regular star forming galaxies than local U/LIRGs (e.g. \cite{rigopoulou_herschel_2014}  \cite{magdis_far-infrared_2014}, \citealp{lee_fine_2017}), whereas other sub-populations include high luminosity starbursts (e.g. \citealp{combes_gas_2013}). This diversity of properties, and sub-populations ranging from normal star forming galaxies to starbursts, ensures intermediate-z U/LIRGs are key to understanding the transition between the two modes of star formation.

Previous studies of the molecular gas content in intermediate-z U/LIRGs have been based either on samples of low luminosity LIRGs (e.g. \citealp{bauermeister_egnog_2013}, \citealp{lee_fine_2017}) or high-luminosity ULIRGs (e.g. \citealp{combes_galaxy_2011}, \citealp{combes_gas_2013}), that do not represent the entire population of U/LIRGs at these redshifts. The main results from these studies show high-luminosity ULIRGs are starbursts with high SFE whereas cold LIRGs lie on, or just above, the MS and are less efficiently forming stars. It is necessary to fill this gap between cold LIRGs and warm ULIRGs if we want to better understand the evolution and properties of U/LIRGs at intermediate-z. 

In this paper we present six targets observed by the NOrthern Ex-tended Millimeter Array (NOEMA) and the Atacama Large Millimeter/submillimeter Array (ALMA). We establish the properties of intermediate-z U/LIRGs, a population of galaxies located in the transition zone between regular star forming galaxies and starbursts. We map the distribution of the molecular gas and dust, measure how efficiently they are forming stars, study their dynamics and investigate what triggers star formation.

When required, values of H$_{0}$ = 70 km s$^{-1}$ Mpc$^{-1}$, $\Omega_{m}$ = 0.3 and $\Omega_{\Lambda}$ = 0.7 were used in this paper. When required the initial mass function (IMF) was assumed to be a \cite{chabrier_galactic_2003} form.

\section{Observations}
\label{section:observations}

\subsection{The Sample}
\label{section:sample}
Our six targets, FLS02 (2 merging galaxies), CDFS1 (2 merging galaxies), SWIRE7 and SWIRE5, were selected from the sample of intermediate redshift U/LIRGs presented in \cite{magdis_far-infrared_2014} (hereafter M14), \cite{rigopoulou_herschel_2014} and \cite{pereira-santaella_optical_2019} (hereafter PS19). Details of the targets are given in Table \ref{tab:sample} and Table \ref{tab:co}. This parent sample was drawn from Herschel deep surveys (\citealp{oliver_herschel_2012}) based on two criteria: having 250 µm fluxes S$_{\rm 250}$ $>$ 150 mJy, and redshifts between 0.2 $<$ z $<$ 0.8. The first criterion ensured the targets would be IR luminous with log(L$_{\rm IR}$) $>$ 11.6. The second criterion was chosen to enable the Spectral and Photometric Imaging REceiver-Fourier Transform Spectrometer (SPIRE-FTS, \citealp{griffin_herschel-spire_2010}) to observe the [CII] 158 $\mu$m emission line. No other criteria on morphologies or colours were applied. The Herschel photometric and spectroscopic observations were complemented with single dish CO measurements (M14), and ground based I-band integral field spectroscopy (PS19). Morphology-wise, PS19 found the parent sample is a mixture of interacting and isolated disks (see Appendix \ref{appendix:opt}). In addition, the sample straddles the region between the MS LIRGs and starbursting ULIRGs at this epoch, so it is ideal for investigating the transition between MS and starburst galaxies and its link to interaction stage, and also enabling a better insight into the U/LIRG population at an epoch when the universe is experiencing a decrease in SFRD. 

For the present study we selected four systems (six galaxies) from the parent sample: two interacting systems (FLS02 and CDFS1) and two systems appearing to be isolated disks (SWIRE5 and SWIRE7) based on optical images with a seeing limited resolution of $\sim$ 2$^{\prime\prime}$ (PS19). To summarise the previous findings, our six targets have dust temperatures ranging between T$_{\rm dust}$ = 29 – 42 K (lower than that of local ULIRGs) and log(L$_{\rm IR}$ / L$_{\odot}$) in the range 11.79 - 12.41. They have similar ratios of un-obscured, traced by H$\alpha$, to obscured, traced by L$_{\rm IR}$, star formation as the local U/LIRGs. They do not exhibit the [CII] / L$_{\rm IR}$ deficiency seen in local ULIRGs, except for FLS02, which is indicative of softer radiation fields and more extended regions of star formation within our sources when compared to local ULIRGs (e.g. \citealp{diaz-santos_herschel/pacs_2017}, \citealp{herrera-camus_shining_2018}).

\begin{table}
	\centering
	\caption{Our Sample of intermediate U/LIRGs}
	\label{tab:sample}
	\begin{threeparttable}
	\begin{tabular}{ lcccc }
        \hline
        Object	&	RA	&	Dec	&	z & Scale  	\\
         & J2000 & J2000 & & kpc arcsec$^{-1}$\\
        \hline
        \hline
        FLS02-N	&	17:13:31.49	&	58:58:04.4	&	0.436 &	5.66 \\
        FLS02-S	&	17:13:31.64	&	58:58:01.0	&	0.437 &	5.65 \\
        CDFS1-W	&	03:29:04.39	&	-28:47:53.0	&	0.289 & 4.34 \\
        CDFS1-E	&	03:29:04.89	&	-28:47:55.5	&	0.291 &	4.36 \\
        SWIRE7	&	11:02:05.68	&	57:57:40.4	&	0.414 &	5.49 \\
        SWIRE5	&	10:35:57.9 	&	58:58:46.2	&	0.366 &	5.08 \\
     
        \hline
        \end{tabular}
    \end{threeparttable}
    
\end{table}

\subsection{NOEMA}
\label{section:noema}
We obtained NOEMA observations of the CO J = 3 $\rightarrow$ 2 $\equiv$ CO(3-2) emission line ($\nu_{\rm rest}$ = 345.796 GHz) and its underlying continuum (rest wavelength $\sim$0.9 mm) for four of our targets (three systems): FLS02, SWIRE5 and SWIRE7 (projects W19BV \& W20GC, PI: S Garcia-Burillo). We used the C configuration of NOEMA and the observations for FLS02 took place over seven nights between 29 December 2019 and 25 February 2020, with SWIRE5 being observed on 28 March 2021 and SWIRE7 on three nights between 31 March 2020 and 04 April 2020. The field of view for these targets was $\approx$ 20-21$^{\prime\prime}$.
The data was reduced using the standard pipeline in the Grenoble Image and Line Data Analysis Software (GILDAS)\footnote{https://www.iram.fr/IRAMFR/GILDAS} software package. Bad visibilities were flagged and removed, and the flux, RF and amplitude/phase were calibrated using standard NOEMA calibrators.
Calibrated uv tables for both continuum and CO line were put together using the Continuum and Line Interferometer Calibration (CLIC)\footnote{https://www.iram.fr/IRAMFR/GILDAS/doc/html/clic-html/clic.html} routine and the uv tables were imaged using the MAPPING routine. The lower side bands were used to put together the continuum uv tables in order to avoid CO line emission. 
We used robust weighting, with a threshold of 1, for the SWIRE7 continuum map, with a natural weighting used for the continuum maps in FLS02 and SWIRE5 in order to maximise the signal to noise ratio (SNR). For the CO(3-2) we used both robust and natural weighting. The natural weighting was used so the CO(3-2) maps have a similar beam size to the continuum maps, allowing the morphology to be compared, with the robust weighting being chosen to improve the spatial resolution when fitting the kinematics (the naturally weighted CO(3-2) cubes are used for all analysis other than fitting the kinematics). The pixel size was chosen to be 0.16$^{\prime\prime}$, which corresponds to $\sim$ 5/6 pixels across the synthesized beam size ($\sim$ 0.8 - 1 $^{\prime\prime}$) for all 3 sources, and we choose a channel width of 20 km s$^{-1}$. 
The dirty cubes were cleaned using the Hogbom algorithm within the CLEAN routine and the fully reduced clean cubes were exported as fits files for further analysis. The final cubes have a resolution of 0.72 - 1.01$^{\prime\prime}$ corresponding to physical scales of $\approx$ 3.7 - 5.7 kpc. Details of integration times, the beam size and noise are given in Table \ref{tab:obs}.

\subsection{ALMA}
\label{section:alma}
We obtained Band 6 ALMA observations of CO(3-2) and continuum for CDFS1 (two merging galaxies) using 43 antennas of the 12-m Array (project 2016.1.00896.S, PI: D Rigopoulou). The observations were taken on 04 November 2016, with a total on source integration time of 0.2 hours and field of view of $\approx$ 23.5$^{\prime\prime}$. One spectral window was centred on the expected frequency of the CO(3-2) line and the other three spectral windows were used to observe the underlying continuum.
The data was reduced using the standard pipeline in the Common Astronomy Software Applications (CASA) software package (\citealp{mcmullin_casa_2007}). Bad visibilities were flagged and removed and we used standard ALMA calibrators to calibrate the source.
The continuum map was created by using multifrequency synthesis mode within CASA on spectral windows that did not contain the CO(3-2) emission line. For the emission line cube we subtracted the continuum emission in each channel by using the UVCONTSUB routine to fit a 0th order polynomial continuum model to the line free channels of the uv plane.
The dirty cubes were then cleaned with the TCLEAN routine and we used natural weighting to maximise the SNR. We choose a pixel size of 0.045$^{\prime\prime}$ which is $\sim$ 7 pixels across the synthesized beam and set the channel width to 20 km s$^{-1}$. All data was corrected for the primary beam response to produce our fully reduced data cubes. The fully reduced data has a resolution of 0.31$^{\prime\prime}$ with a physical scale $\approx$ 1.5 kpc.

\begin{table}
	\centering
	\caption{Observation details for our sample.}
	\label{tab:obs}
	\begin{threeparttable}
	\begin{tabular}{ lccc }
        \hline
        Object & Integration time\tnote{1} & Synthesized Beam\tnote{2}	 & $\sigma_{\rm rms}$\tnote{3} \\
        & hours & \arcsec $\times$ \arcsec & mJy/beam  \\
        \hline\hline
        CDFS1	 & 0.2 & (0.31 $\times$ 0.28)\tnote{a}	& 0.91 \\
        FLS02	 & 9.8 & 0.82 $\times$ 0.67\tnote{a} &  0.55	 \\
        & & (1.01 $\times$ 0.78) & \\
        SWIRE5	& 3.8 & 0.72 $\times$ 0.61  & 0.61	\\
        & & (0.82 $\times$ 0.67)  &\\
        SWIRE7	& 7.1 & 0.90 $\times$ 0.79 & 0.71 \\
        
        \hline
    \end{tabular}
    \begin{tablenotes}
        \item[1] On source integration time.
        \item[2] Synthesized beam for the fully reduced data.
        \item[3] 1$\sigma$ sensitivity in each 20 km s$^{-1}$ channel.
        \item[a]  Value in brackets is the synthesized beam for natural weighting. Values without brackets are the synthesized beam for robust weighting with a threshold of 1. See Section \ref{section:noema} for full details.
    \end{tablenotes}
    \end{threeparttable}
\end{table}

\section{Analysis}
\label{section:analysis}

\subsection{CO(3-2) and Continuum Emission}
\label{section:maps}

\begin{figure*}
    \centering
    \begin{subfigure}{0.5\textwidth}
        \includegraphics[width=\linewidth]{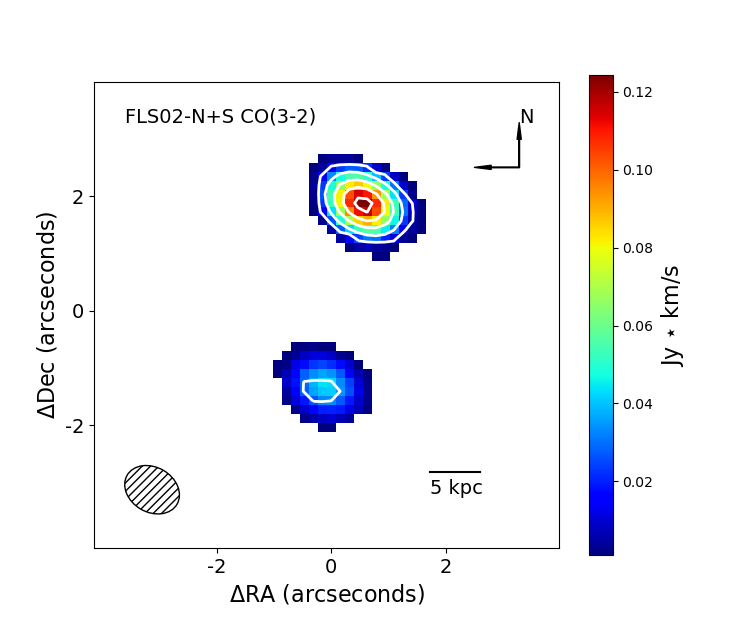}
    \end{subfigure}\hfil
    \begin{subfigure}{0.5\textwidth}
        \includegraphics[width=\linewidth]{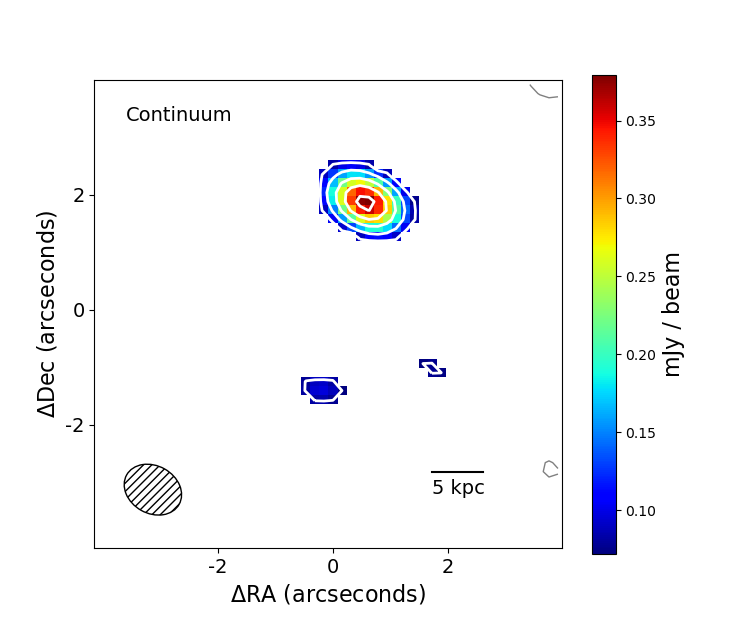}
    \end{subfigure}
    
    \centering
    \begin{subfigure}{0.5\textwidth}
        \includegraphics[width=\linewidth]{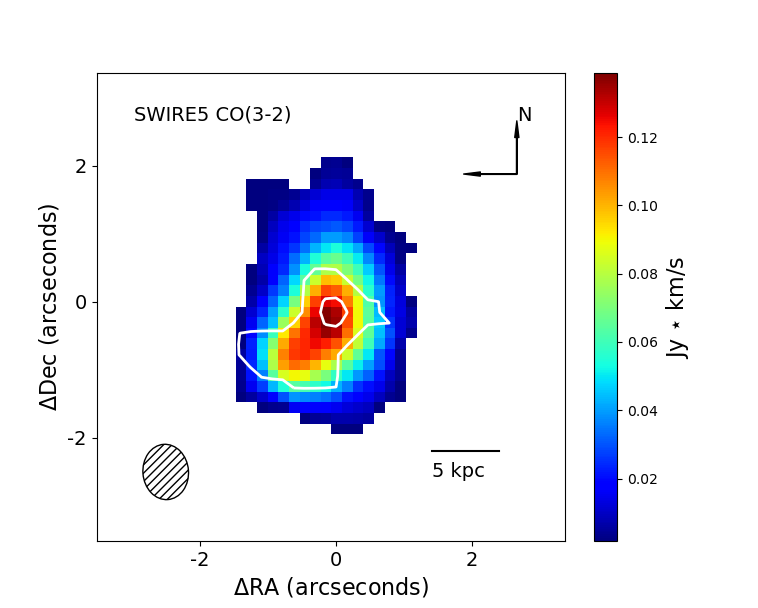}
    \end{subfigure}\hfil
    \begin{subfigure}{0.5\textwidth}
        \includegraphics[width=\linewidth]{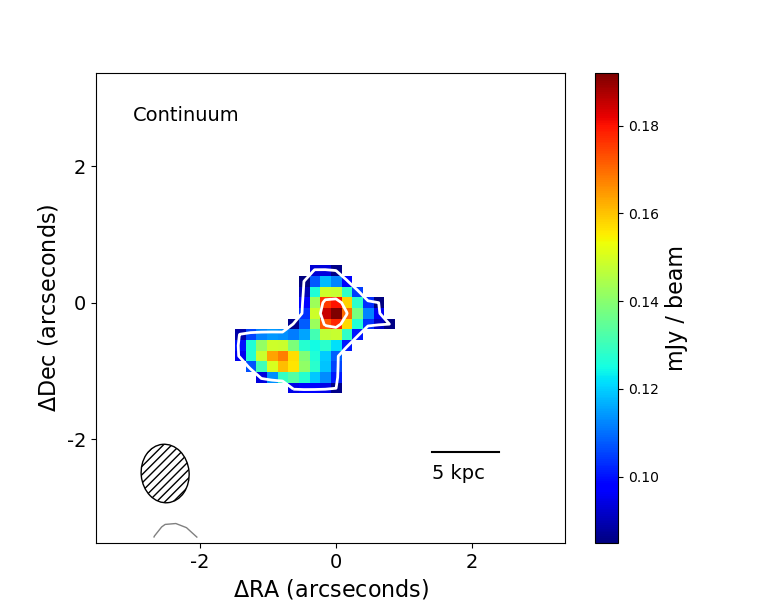}
    \end{subfigure}
    
    \centering
    \begin{subfigure}{0.5\textwidth}
        \includegraphics[width=\linewidth]{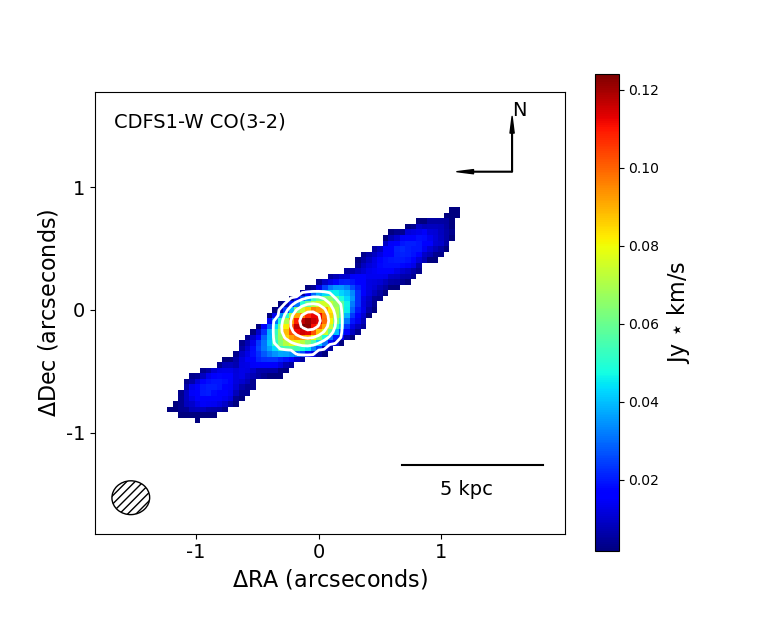}
    \end{subfigure}\hfil
    \begin{subfigure}{0.5\textwidth}
        \includegraphics[width=\linewidth]{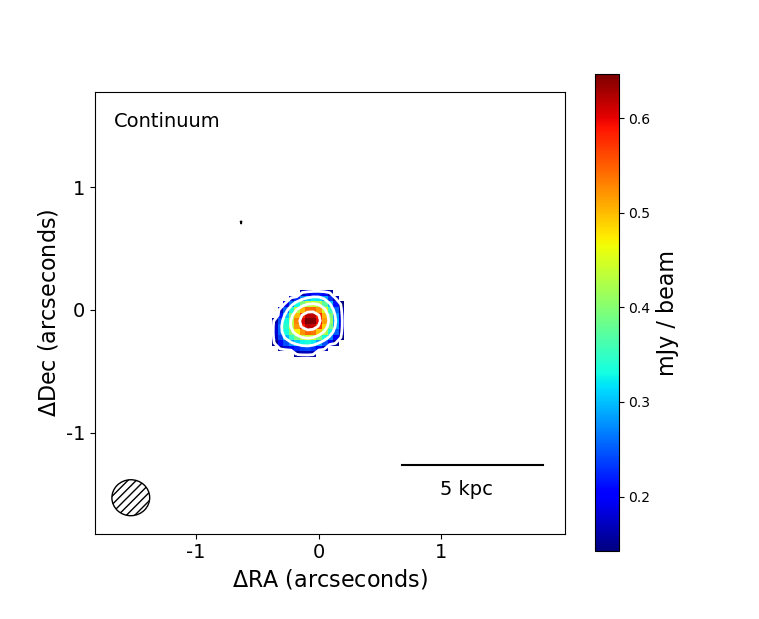}
    \end{subfigure}
    
    \caption{Intensity maps for our targets with CO(3-2), in units of Jy km s$^{-1}$, on the left and continuum (rest wavelength $\sim$ 0.9mm), in units of mJy/beam, on the right. The CO(3-2) maps were made using a 3$\sigma$ cut in each channel and a 3$\sigma$ cut was also used on the continuum maps. Contours of the continuum emission are overlaid in white on the CO(3-2) maps to show the overlap of dust and molecular gas, with the contours representing multiples of 3$\sigma$ continuum emission (i.e. 3$\sigma$, 6$\sigma$, 9$\sigma$ etc.). The grey contours on the continuum plots show negative 3$\sigma$ emission. The beam size is shown in the bottom left corner of each plot and the physical size scale is shown on the bottom right. For optical maps of the same galaxies see Appendix \ref{appendix:opt} and PS19.} 
    \label{fig:maps}
\end{figure*}

\begin{table*}
	\centering
	\caption{Observed CO(3-2) properties of our sample of intermediate-z U/LIRGs.}
	\label{tab:co}
	\begin{threeparttable}
	\begin{tabular}{ lcccccccc }
        \hline
        Object & S$_{\rm CO}\Delta$v (total)\tnote{1} & L'$_{\rm CO}$ (total)\tnote{2} &	S$_{\rm CO}\Delta$v (masked)\tnote{3} & L'$_{\rm CO}$ (masked)\tnote{4} & log(L$_{\rm IR}$ / L$_{\odot}$)\tnote{5}	& T$_{\rm dust}$\tnote{6}  & SFR$_{\rm IR}$\tnote{7}  & 12+log(O/H)\tnote{8}  \\
        &  Jy km s$^{-1}$  & 10$^{9}
        $ K km s$^{-1}$ pc$^{2}$ & Jy km s$^{-1}$  & 10$^{9}
        $ K km s$^{-1}$ pc$^{2}$ &  & K & M$_{\odot}$ yr$^{-1}$ &  \\
        \hline\hline
        CDFS1-W	 & 13.42 $\pm$ 1.11 & 6.26 $\pm$ 0.52 & 9.71 $\pm$ 0.66 & 4.53 $\pm$ 0.31 & 11.79 & 29 & 91 &  9.08	\\
        FLS02-N	& 4.50 $\pm$ 0.33 & 4.93 $\pm$ 0.36 & 4.39 $\pm$ 0.29 & 4.81 $\pm$ 0.32 & 12.41\tnote{a} & 42\tnote{a} & 353 &	8.87	\\
        FLS02-S	&  1.34  $\pm$ 0.18 & 1.47 $\pm$ 0.20 & 1.34  $\pm$ 0.18\tnote{b} & 1.47 $\pm$ 0.20\tnote{b} & - & -	& 27 &	8.82 \\
        SWIRE5	& 13.36 $\pm$ 0.87  & 10.18 $\pm$ 0.66 & 8.94 $\pm$ 0.56 & 6.81 $\pm$ 0.43	& 12.06 & 33 & 170 &  8.77\\
        SWIRE7	& $<$0.26\tnote{c} & $<$0.26\tnote{c} &  $<$0.26 & $<$0.26  & 12.10	& 36 & 190 & - \\
        \hline
    \end{tabular}
    \begin{tablenotes}
        \item[1] The total CO(3-2) flux density of the galaxy.
        \item[2] The total CO(3-2) line luminosity of the galaxy.
        \item[3] The flux density within the masked region of each galaxy as explained in Section \ref{section:gas}.
        \item[4] The CO(3-2) line luminosity within the masked region.
        \item[5] The IR luminosity from \cite{magdis_far-infrared_2014}.
        \item[6] The dust temperature from \cite{magdis_far-infrared_2014}.
        \item[7] The IR star formation rate from \cite{pereira-santaella_optical_2019}.
        \item[8] The gas phase oxygen abundance obtained from the N2 ratio presented in \cite{pereira-santaella_optical_2019}.
        \item[a] The log(L$_{\rm IR}$ / L$_{\odot}$) of 12.41 and T$_{\rm dust}$ of 42 K is for the total FLS02 system. The decomposition of the L$\rm IR$ into SFR for the individual galaxies of FLS02 is explained in Section \ref{section:sfe}.
        \item[b] Masked region is equal to the total CO(3-2) region for FLS02-S as explained in Section \ref{section:sfe}.
        \item[c] SWIRE7 undetected in CO(3-2) so the upper limit is calculated from the masked region.
    \end{tablenotes}
    \end{threeparttable}
\end{table*}

\begin{figure}
    \centering
    \includegraphics[scale = 0.5]{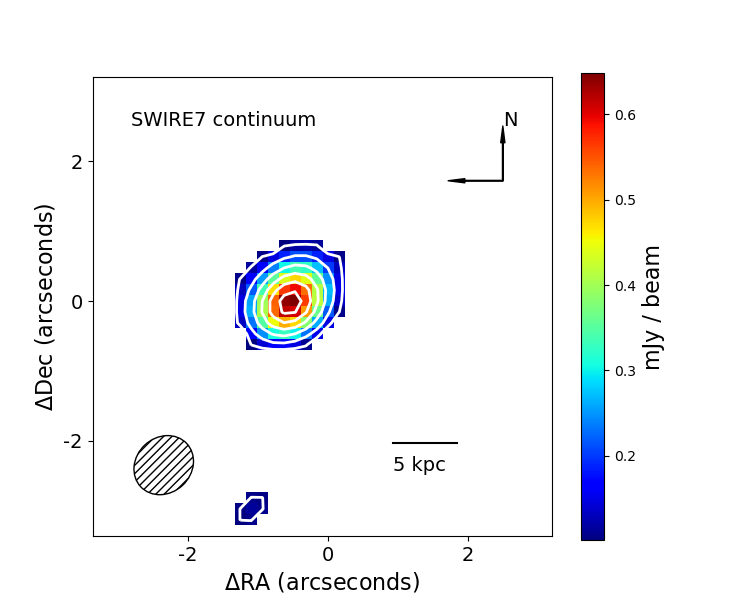}
    \caption{Same as Figure \ref{fig:maps} except for SWIRE7 continuum. No CO(3-2) was observed in this target.}
    \label{fig:cont_SWIRE7}
\end{figure}

We created intensity maps of both CO(3-2) and continuum for our sources to study their morphology, flux and relative positions. To make the CO(3-2) intensity maps we performed a 3$\sigma$ cut in each channel of the data cube and then integrated the remaining flux across the total width of the emission line. For the continuum 0th moment maps we made a 3$\sigma$ cut on the continuum map. These plots are presented in Figures \ref{fig:maps} and \ref{fig:cont_SWIRE7}, with 3$\sigma$ (and multiples of 3$\sigma$) contours of the continuum emission overlaid on the CO(3-2) maps. Line widths are in agreement with previous measurements using a single dish observation in M14. The galaxy integrated spectra showing complex dynamics (see Appendix \ref{appendix:spectra}) and we only use the spatially resolved NOEMA and ALMA data throughout this study. 

In this analysis we assume that the main source of the continuum emission, observed at a rest wavelength $\sim$ 0.9 mm, arises from the Rayleigh-Jeans tail of the dust emission. It is difficult to rule out AGN contribution, but as our sample do not show broad components in the H$\alpha$ emission line (PS19) we assume they are star formation dominated. The absorption cross section of galactic dust strongly peaks in the UV so the IR emission is attributable to dust re-processed stellar light from underlying O-type and B-type stars. This makes L$_{\rm IR}$ emission a probe of star formation within the last 100 Myr (\citealp{kennicutt_star_2012}). The CO(3-2) transition is 33 K above the ground state and has a critical density of $>$ 10$^{4}$ cm$^{-3}$ at temperatures $<$ 20 K, so it traces the molecular clouds within the ISM from which stars are formed (e.g. \citealp{rigby_chimps_2016}), hence these observations allow us to explore the relationship between molecular gas and ongoing star formation.

For the northern source in FLS02 (hereafter FLS02-N) the continuum and CO(3-2) emission are approximately co-spatial and the centre of the continuum coincides with the centre of CO(3-2). The southern source, hereafter FLS02-S, has weaker CO(3-2) emission with a marginal $\sim$ 4$\sigma$ detection in the continuum (see Figure \ref{fig:maps}). The SWIRE5 CO(3-2) and continuum emission show very different morphology with the star forming region smaller in extent than the molecular gas and appears to have two peaks which are separated by $\sim$ 7 kpc (see Figure \ref{fig:maps}). In CDFS1, the western source, CDFS1-W, shows a disk like structure in CO(3-2) with a more compact continuum located in the central region of the disk (see Figure \ref{fig:maps}).
Neither CO(3-2) or continuum are observed in CDFS1-E despite this galaxy having a H$\alpha$ detection in PS19 (see Appendix \ref{appendix:companion} for its expected position with respect to CDFS1-W and Figure \ref{fig:CDFS1_optical} for the optical observations of this system, noting that north points down in the optical image). No CO(3-2) was detected in SWIRE7, but there is a 20$\sigma$ detection of the continuum (see Figure \ref{fig:cont_SWIRE7}). In Appendix \ref{appendix:opt} we have included the optical images of all four systems from PS19 for comparison with our CO(3-2) and sub-mm continuum. The morphology and its link to both the triggers of star formation and the SFE within a galaxy will be discussed in more detail in Section \ref{section:triggers}. 

\subsection{Molecular Gas Mass}
\label{section:gas}

\begin{figure*}
  \centering
  \includegraphics[scale=0.5]{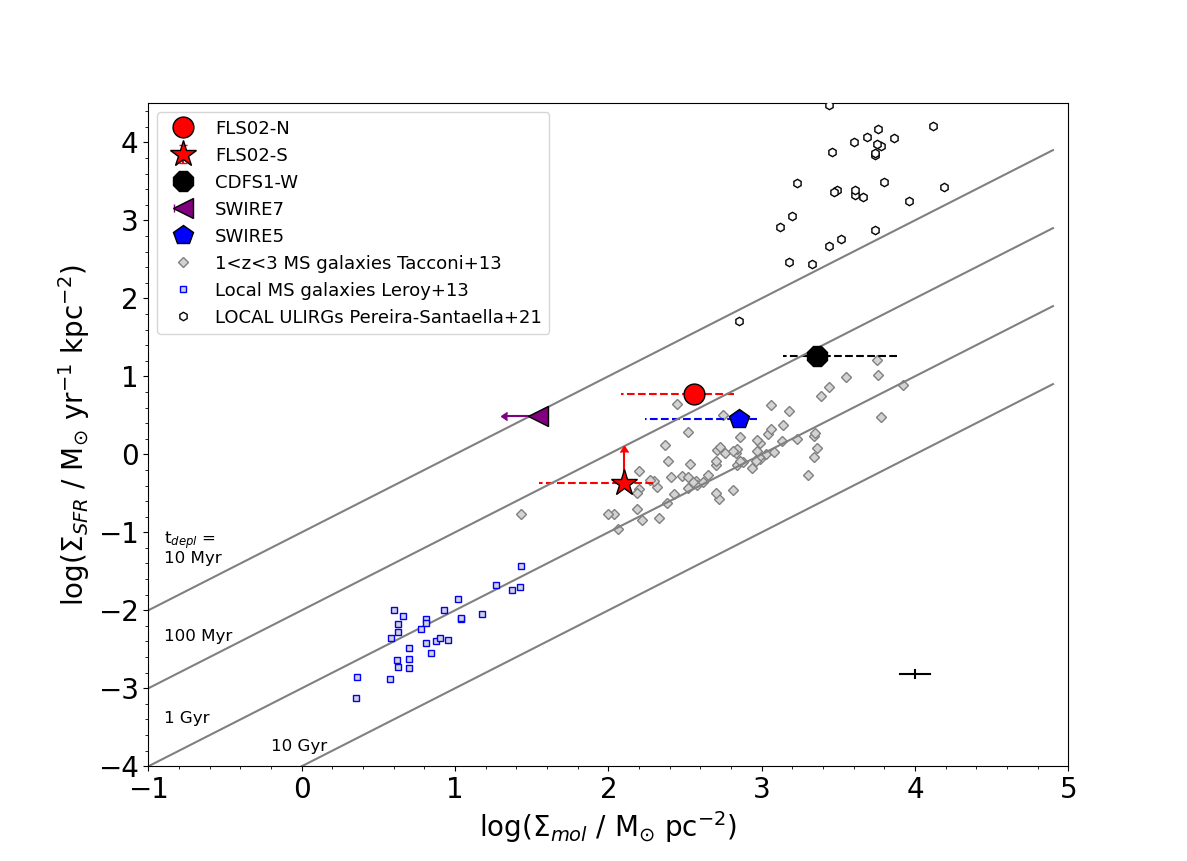}
  \caption{Gas surface-density, $\Sigma_{\rm mol}$, versus star-formation surface-density, $\Sigma_{\rm SFR}$, for our sample of U/LIRGs. They are compared to values for high-z star forming galaxies (grey diamonds, \citealp{tacconi_phibss:_2013}), local ULIRGs (white hexagons, \citealp{pereira-santaella_are_2021}) and local star forming disk galaxies (blue squares, \citealp{leroy_molecular_2013}). It can be seen our sample of U/LIRGs have longer depletion times when compared to local ULIRGs, but shorter depletion times when compared to normal star forming galaxies, showing our intermediate-z U/LIRGs falls in the transition region between starburst and main sequence galaxies. The dashed lines show how changing the $\alpha_{\rm CO}$ affects the $\Sigma_{\rm mol}$ taking the extreme $\alpha_{\rm CO(1-0)}$ values: the local ULIRG conversion factor of 0.8 M$_{\odot}$ / (K km s$^{-1}$ pc$^{2}$) and the Milky Way value of 4.36 M$_{\odot}$ / (K km s$^{-1}$ pc$^{2}$). The cross in the bottom right corner shows the typical size of the error bars.}
  \label{fig:ks}
\end{figure*}

Local observations have shown the integrated flux from the low-J CO transitions can be used to estimate the molecular gas mass within a galaxy (e.g. \citealp{bolatto_co--h2_2013} and references therein). Firstly, we must convert the CO(3-2) flux density from our intensity maps to the integrated line luminosity using the following equation from \cite{solommon_molecular_1997}, 
\begin{equation}
L'_{\rm{CO(J \; - \; J-1)}} = 3.25 \times 10^{7} \cdot S_{\rm CO}\Delta \rm{v} \cdot \nu^{-2}_{\rm obs} \cdot D_{\rm L}^{2} \cdot (1+z)^{-3}      
\end{equation}
where S$_{\rm CO}\Delta$v is the integrated flux density in Jy km s$^{-1}$, $\nu_{\rm obs}$ is the observed frequency in GHz, $D_{\rm L}$ is the luminosity distance in Mpc, z is the redshift of the source and the line luminosity is in units of K km s$^{-1}$ pc$^{2}$.

The line luminosity is proportional to the gas within a galaxy, therefore in order to estimate the molecular gas mass, M$_{\rm mol}$, from L$^{\prime}_{\rm{CO(3-2)}}$ the use of an appropriate $\alpha_{\rm CO}$ conversion factor is required. This factor is dependent on the density of molecular clouds, the Rayleigh-Jeans brightness temperature and inversely proportional to the metallicity within a galaxy (see \citealp{bolatto_co--h2_2013} for an in-depth review). To account for metallicity we adopt a metallicity dependent conversion factor described in \cite{genzel_metallicity_2012} and \cite{genzel_combined_2015}:
\begin{equation}
\alpha_{\rm CO (3-2)}(Z) \; = \; \alpha_{\rm 0} \cdot \chi(Z) \cdot R_{1-3} \quad (\rm{M_{\odot} / (K  \; km \; s^{-1} \; pc^{2})})   
\label{eqt:alpha}
\end{equation}
where $\alpha_{\rm 0}$ is the un-scaled J = 1 $\rightarrow$ 0 conversion factor, $\chi$(Z) is the metallicity dependent scaling factor and R$_{1-3}$ is the ratio of the L$^{\prime}_{\rm CO(1-0)}$ to L$^{\prime}_{\rm CO(3-2)}$ which is required to correct for the lower Rayleigh-Jeans brightness temperatures in higher J transitions. We choose the metallicity dependent scaling factor of  \cite{genzel_combined_2015}):
\begin{equation}
\chi (Z) = 10^{-1.27 \times (\rm{12 + log(O/H)} - 8.67)}    
\label{eqt:scaling}
\end{equation}
where 12 + log(O/H) is the gas phase oxygen abundance. The constant 8.67 ensures Equation \ref{eqt:alpha} reverts to the $\alpha_{\rm MW}$ when the gas phase oxygen abundance equals the solar abundance. 

We assume $\alpha_{\rm 0}$ equals the Milky Way conversion factor $\alpha_{\rm MW}$ = 4.36 $\pm$ 0.9 M$_{\odot}$ / (K km s$^{-1}$ pc$^{2}$), including the correction for Helium, and R$_{1-3}$ = 1.9 (\citealp{lamperti_co32co10_2020}), which is a typical value seen in star forming galaxies. For the metallicity of our sample, we use our sample's [NII] 6583 \AA\ / H$\alpha$ ratios (N2) reported in PS19 and converted these to 12 + log(O/H) using the \citealp{pettini_[o_2004} correlation. For SWIRE7 we set $\chi$(Z) = 1 as it had no [NII] 6583 \AA\ detection in PS19. The values we obtain for $\alpha_{\rm{CO (3-2)}}(Z)$ are shown in Table \ref{tab:ks}. Thus, the molecular gas mass in each target is found by 
\begin{equation}
M_{\rm mol} = \alpha_{\rm{CO (3-2)}}(Z) \cdot L'_{\rm{CO(3-2)}}
\end{equation}

One caveat when choosing a conversion factor is that local ULIRGs typically have an $\alpha_{\rm CO}$ factor which is about five times lower than that of the Milky Way, with a value of $\approx$ 0.8 M$_{\odot}$ / (K km s$^{-1}$ pc$^{2}$) for the CO(1-0) transition (e.g. \citealp{downes_rotating_1998}). Our sample of U/LIRGs have different ISM conditions compared to those of local ULIRGs (as discussed in M14) and are in an earlier stage of interaction (PS19). Therefore the metallicity dependent $\alpha_{\rm{CO (3-2)}}$(Z) is the most appropriate conversion factor for our sample. We discuss how the lower $\alpha_{\rm CO}$ would affect our results in Section \ref{section:gasandsfr}. 

It is worth emphasising that our choice of metallicity scaling factor is quite conservative. \cite{genzel_combined_2015} uses two metallicity scaling functions (see equations 6 and 7 of \citealp{genzel_combined_2015}) and takes the geometric mean of both as the final scaling factor. We chose Equation \ref{eqt:scaling} to be our scaling factor as it gave us the lowest $\alpha_{\rm{CO (3-2)}}(Z)$. If we chose the geometric mean, as in \citealp{genzel_combined_2015}, then our $\alpha_{\rm{CO (3-2)}}(Z)$ would be $\sim$ 15\% higher on average, with a corresponding increase in M$_{\rm mol}$.

Finally, M14 estimated the M$_{\rm mol}$ using the dust to gas mass ratio technique (see M14 and references therein) and inferred an $\alpha_{\rm CO}$ consistent with local star forming galaxies for our sample (with the exception of FLS02), further corroborating our choice. We note the exception of FLS02 in M14 was due to its observed metal deficiency, based on the [CII] emission line, and classification as an AGN. Subsequent evidence, such as a narrow H$\alpha$ and [NII] 6583 \AA\ emission lines and polycyclic aromatic hydrocarbons features consistent with mixed objects (Infrared Spectrograph data from \citealp{houck_spitzer_2007}), suggest that the source is not an AGN dominated object. Hence, our scaled $\alpha_{\rm CO}$(Z) is the most pertinent choice for FLS02.

\subsection{Star Formation Efficiency}
\label{section:sfe}

A galaxy's star forming efficiency is calculated from its surface density of SFR and molecular gas (SFE = $\Sigma_{\rm SFR} / \Sigma_{\rm mol}$). To get the surface density, we first found the spatial size of the star forming region, A$_{\rm cont}$, within our U/LIRGs by assuming that the continuum flux arises from the dust and that it traces the full extent of obscured star formation within the galaxy (SFR$_{\rm IR}$). As our sources are resolved we measured this area directly from the moment maps by multiplying the area of each pixel by the number of pixels with SNR > 3$\sigma$ (see Figure \ref{fig:maps}). To determine the SFR in each of our targets we take the L$_{\rm IR}$-SFR$_{\rm IR}$ conversion factor in \citealp{murphy_calibrating_2011} and apply this to the L$_{\rm IR}$ from M14. As our sources are heavily dust obscured (PS19) the SFR$_{\rm IR}$ $\approx$ total SFR. 

As discussed in Section \ref{section:maps} the morphology of the continuum and CO emission is different across our sample. To accurately determine the $\tau_{\rm dep}$ we use only CO(3-2) that overlaps with the SFR$_{\rm IR}$. To do this we placed a mask on the emission line cube that ensured only CO(3-2) flux that arises within A$_{\rm cont}$ is extracted. This mask corresponds to the outer white contour in Figure \ref{fig:maps}. All flux outside this contour is not used for the SFE calculation. To integrate the masked CO(3-2) flux we narrowed the velocity window to include only channels containing line emission and applied no threshold within the masked region to ensure we missed no flux. The surface density for the molecular gas was then calculated using $\Sigma_{\rm mol}$ = M$_{\rm mol}$ / A$_{\rm cont}$ and the SFR surface density was found via $\Sigma_{\rm SFR}$ = SFR$_{\rm IR}$ / A$_{\rm cont}$, where A$_{\rm cont}$ is the area of the star forming region as defined at the start of this section.

We estimated an upper limit for M$_{\rm mol}$ in SWIRE7 by finding a 3$\sigma$ upper limit for S$_{\rm CO}\Delta$v. This was calculated from 3$\sigma$/beam $= 3\sqrt{N} \delta v \sigma_{\rm rms}$, where N is the number of channels, $\delta v$ is the channel width in km s$^{-1}$ and $\sigma_{\rm rms}$ is the noise per channel. We assumed a linewidth of 300 km s$^{-1}$ and multiplied the 3$\sigma$/beam by the square root of the number of beams within A$_{\rm cont}$ of SWIRE7 to get the upper limit. For FLS02-S there is a marginal ($\approx$ 4$\sigma$) unresolved continuum detection so we assumed A$_{\rm cont}$ is equal in size to the CO(3-2) emitting region.

Our results are listed in Tables \ref{tab:co} and \ref{tab:ks}, and plotted in Figure \ref{fig:ks}. Our sample of U/LIRGs show a $\tau_{\rm dep}$ of 11 - 301 Myr, 
with $\tau_{\rm dep}$ of 62 - 301 Myr, excluding SWIRE7 as the source was not detected in the CO(3-2) emission line (throughout this paper we use SFE and $\tau_{\rm dep}$ interchangeably with a higher SFE equalling a lower $\tau_{\rm dep}$ and vice versa for a low SFE). This is higher than the $\tau_{\rm dep}$ = 1 - 15 Myr found for local U/LIRGs (\citealp{pereira-santaella_are_2021}), but lower than the values found for regular star forming population at both low and high redshifts which is on the order of 1 Gyr (e.g. \citealp{leroy_molecular_2013}, \citealp{tacconi_phibss:_2013}). This demonstrates that our U/LIRGs, which fall within the transition region between starbursts and MS galaxies also fall between the loci of major mergers and regular star forming galaxies on the Kennicutt-Schmitt relation. Figure \ref{fig:ks} also shows how the $\Sigma_{\rm mol}$ values of our intermediate-z U/LIRGs are similar to those of gas rich MS galaxies at high-z. However,  our intermediate-s U/LIRGs have a higher SFE as shown by our sample having, on average, a larger $\Sigma_{\rm SFR}$ for a given $\Sigma_{\rm mol}$. 

It is worth noting that the SFR$_{\rm IR}$ has been computed from the L$_{\rm IR}$ and the latter was obtained by fitting the photometry from Spitzer and Herschel Space Telescopes. We are, therefore, constrained by the resolution afforded by the Herschel$/$SPIRE bands, which is $\approx$ 20 arcseconds (\citealp{griffin_herschel-spire_2010}). Hence, the L$_{\rm IR}$ reported for the two interacting systems, FLS02 (centres separated by $\approx$ 4$^{\prime \prime}$) and CDFS1 (separated by $\approx$ 7$^{\prime \prime}$), corresponds to the entire system and, cannot be calculated for each of the two nuclei independently. Since CDFS1-E is not observed in either continuum or CO(3-2), it is not included in this analysis, and we assume that all the SFR$_{\rm IR}$ arises from CDFS1-W. For FLS02 we measure the total continuum flux from both sources and find each galaxy's relative contribution to the total flux. This apportions the SFR$_{\rm IR}$ = 380 M$_{\odot}$ yr$^{-1}$ for the system into 353 M$_{\odot}$ yr$^{-1}$ for FLS02-N and 27 M$_{\odot}$ yr$^{-1}$ for FLS02-S.

Our motivation for using the A$_{\rm cont}$, i.e. the 3$\sigma$ cut on the continuum, to calculate the surface densities, is to ensure maximum overlap between star forming regions and molecular gas. If instead we assume a larger area which includes regions with CO(3-2) flux but no continuum emission, our average $\tau_{\rm dep}$ increases. This is due to the CO(3-2) emitting region being larger than the continuum emitting region, as seen in Figure \ref{fig:maps}. Therefore $\Sigma_{\rm SFR}$ decreases faster than $\Sigma_{\rm mol}$ as the SFR$_{\rm IR}$ stays constant when we make the area greater than A$_{\rm cont}$ (we assumed all SFR$_{\rm IR}$ arises from the 3$\sigma$ continuum cut), whereas we increase the measured CO(3-2) flux, even if each additional pixel has a lower surface density than the average we calculated in Table \ref{tab:co}. Depending on how we select this larger region, e.g. a 3$\sigma$ or 5$\sigma$ cut on the CO(3-2) maps (which is equivalent to choosing  a lower $\sigma$ threshold on the continuum maps), the average $\tau_{\rm dep}$ of our sample only increases by at most 40 Myr and therefore our conclusions remain unchanged. Our sample is still far below the typical $\tau_{\rm dep}$ of 1 Gyr in normal star forming galaxies (e.g. \citealp{leroy_molecular_2013}, \citealp{tacconi_phibss:_2013}). 

Similarly, the (de)convolved source size also does not affect our results, as the beam sizes for the continuum and line emission are similar, and any CO(3-2) flux smeared outside A$_{\rm cont}$ will only marginally change the $\Sigma_{\rm mol}$ and $\tau_{\rm dep}$. As a test for this we assumed that the total integrated galaxy flux (see column 3 in Table \ref{tab:co}) arises from within A$_{\rm cont}$ and found that the increase in $\Sigma_{\rm mol}$ changes the average $\tau_{\rm dep}$ by less than 40 Myr, demonstrating the robustness of our findings.

\begin{table*}
	\centering
	\caption{Surface density details for our sample of intermediate-z U/LIRGs}
	\label{tab:ks}
	\begin{threeparttable}
	\begin{tabular}{ lccccccc }
        \hline
        Object & $\alpha_{\rm CO(3-2)}$(Z)\tnote{1} & M$_{\rm mol}$ (total)\tnote{2} & M$_{\rm mol}$ (masked)\tnote{3} &  A$_{\rm cont}$\tnote{4} & log($\Sigma_{\rm CO}$ / M$_{\odot}$ pc$^{-2}$)\tnote{5} & log($\Sigma_{\rm SFR}$ / M$_{\odot}$ yr$^{-1}$ kpc$^{-2}$)\tnote{6} & $\tau_{\rm dep}$\tnote{7}  \\
         &  M$_{\odot}$ / (K  km s$^{-1}$ pc$^{2}$)  & 10$^{10}$ M$_{\odot}$ & 10$^{10}$ M$_{\odot}$ & 10$^{7}$ pc$^{2}$  &  &  & Myr \\
        \hline\hline
        CDFS01-W	&	2.51	$\pm$ 0.52	& 1.57 $\pm$ 0.35	&	1.14	$\pm$ 0.25		&	0.48	&	3.37	$\pm$	0.09	&	1.27	&	125 $\pm$ 30 \\
        FLS02-N	&	4.56	$\pm$	0.94 & 2.25 $\pm$ 0.49	&	2.20	$\pm$ 0.48		&	6.04	&	2.56	$\pm$	0.09	&	0.77		&	62 $\pm$ 15 \\
        FLS02-S	&	5.42	$\pm$	1.12 & 0.80	$\pm$	0.20\tnote{a}	&	0.80	$\pm$	0.20	&	6.35\tnote{a}	&	2.10	$\pm$	0.1	&	-0.38		&	301 $\pm$ 80 \\
        SWIRE 5	&	6.27	$\pm$	1.29  & 6.38 $\pm$ 1.38	&	4.27	$ \pm$ 0.92		&	6.02	&	2.85	$\pm$	0.09	&	0.45		&	251 $\pm$ 60 \\
        SWIRE7	&	8.28	$\pm$ 0.18	& $<$0.21\tnote{b}	&	$<$0.21		&	6.12	&	$<$1.54		&	0.49		&	$<$ 11 \\

        \hline
    \end{tabular}
    \begin{tablenotes}
        \item[1] The conversion factor from L$'_{\rm{CO(3-2)}}$ to molecular gas mass as described in Equation \ref{eqt:alpha}. This value includes a factor of 1.9 to go from L$'_{\rm{CO(3-2)}}$ to L$'_{\rm{CO(1-0)}}$.
        \item[2] The total molecular gas mass within the galaxy 
        \item[3] The molecular gas mass inside the masked region as described in Section \ref{section:gas} 
        \item[4] The area of the continuum, assuming a 3$\sigma$ cut.
        \item[5] The molecular gas mass surface density.
        \item[6] The star formation rate surface density. We assume an error of 10\% for each target.
        \item[7] The depletion time of the molecular gas reservoirs within our galaxies.
        \item[a] The masked region is equal to the total CO(3-2) region for FLS02-S as explained in Section \ref{section:sfe}.
        \item[b] SWIRE7 is undetected in CO(3-2) so the upper limit is calculated from the masked region as explained in Section \ref{section:sfe}.
    \end{tablenotes}
    \end{threeparttable}
\end{table*}

\begin{figure*}
  \centering
  \includegraphics[scale=0.5]{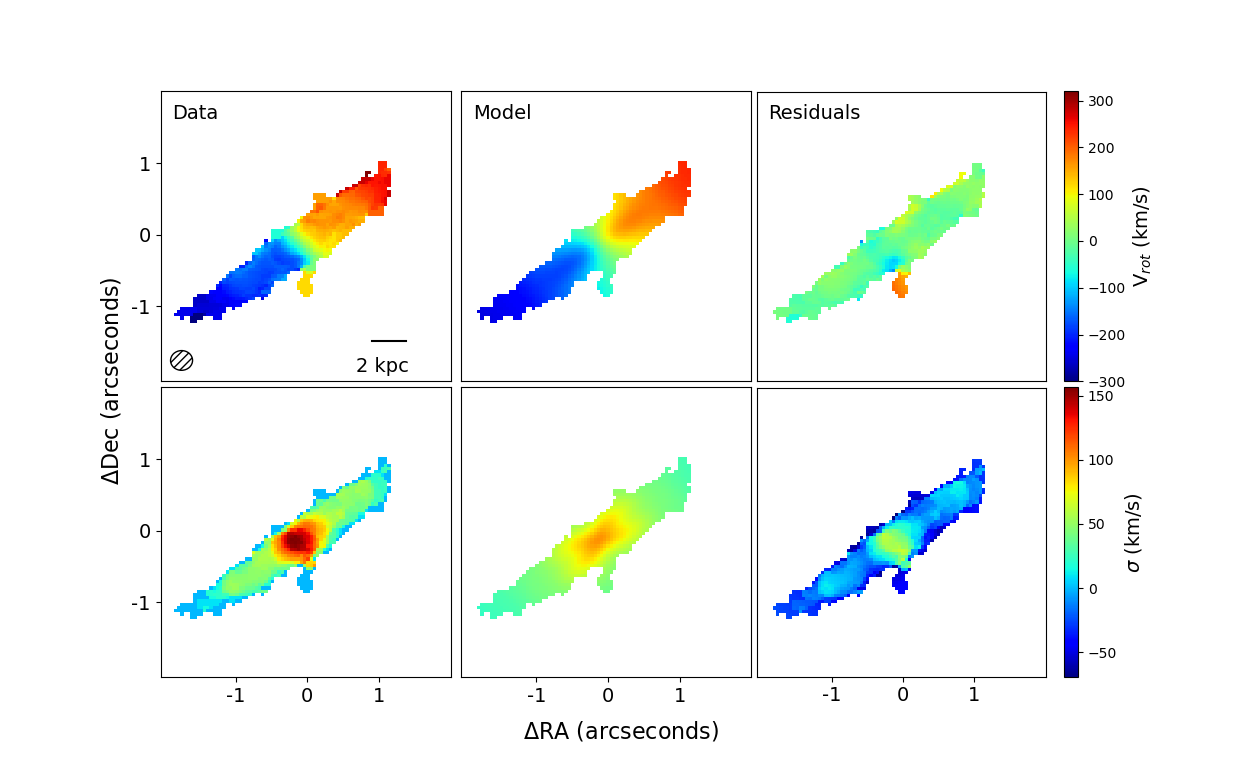}
  \caption{Moment maps from the observations and best-fit rotating disk model, found by $^{\rm 3D}$BAROLO, for CDFS1-W. The top left image shows the observed 1st moment map, the top middle image is the 1st moment map from the best-fit model and the top right is the residual map made by subtracting the best-fit model from the observed map. The bottom left shows the observed 2nd moment map, bottom middle the best-fit 2nd moment map and the bottom right is the residual map. The beam size and physical scale bar are shown in the top left plot. Details of the kinematic modelling are given in Section \ref{section:kinematics}.}
  \label{fig:CDFS1}
\end{figure*}

\begin{figure*}
  \centering
  \includegraphics[scale=0.45]{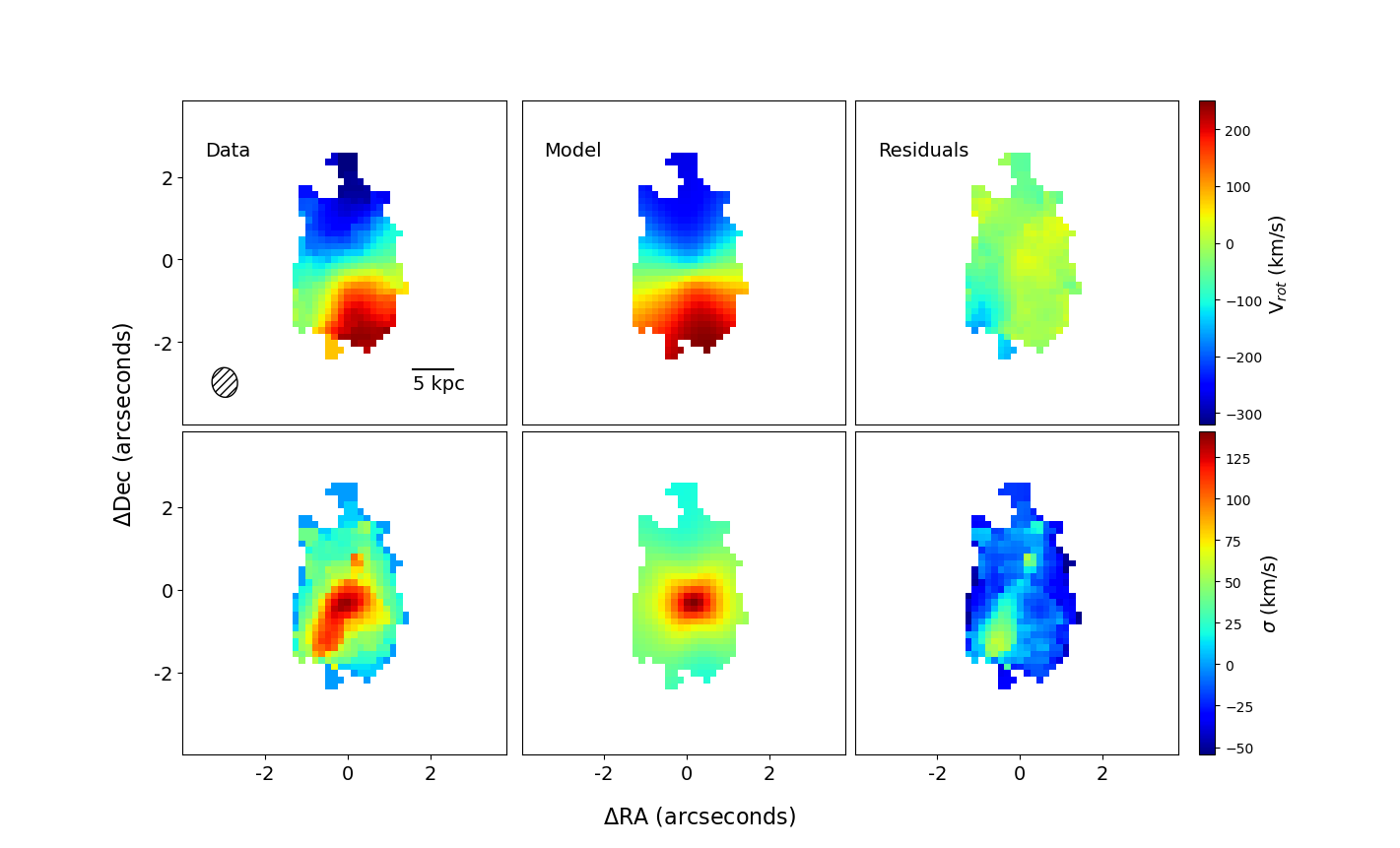}
  \caption{Details the same as in Figure \ref{fig:CDFS1}, but for SWIRE5.}
  \label{fig:SWIRE5}
\end{figure*}

\begin{figure*}
    \begin{subfigure}{0.5\textwidth}
        \centering
         \includegraphics[width=0.65\linewidth]{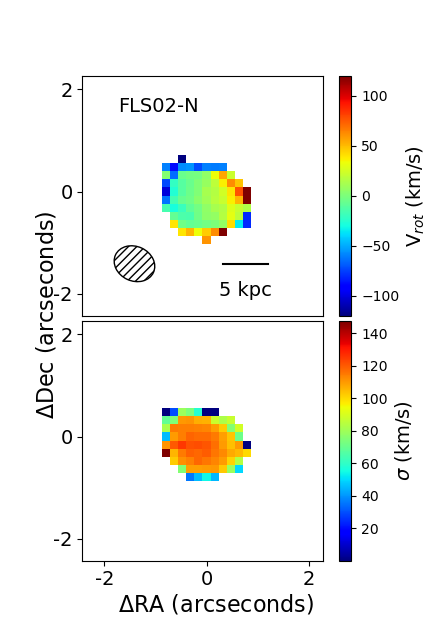}
    \end{subfigure}\hfil
    \begin{subfigure}{0.5\textwidth}
        \centering
         \includegraphics[width=0.65\linewidth]{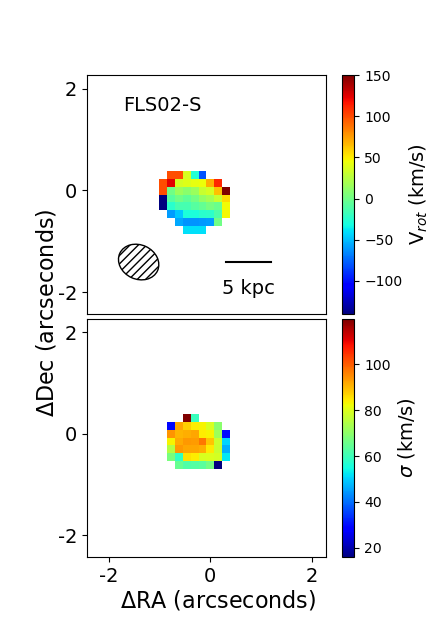}
    \end{subfigure}\hfil
    \caption{ The top panels shows the 1st moment maps and bottom panels show the 2nd moment maps for FLS02-N (left) and FLS02-S (right). A $^{\rm 3D}$BAROLO model disk could not be adequately fitted to this data due to insufficient resolution. Both galaxies show hints of a velocity structure consistent with rotating disk but we need higher resolution imaging, to resolve the velocity dispersion and to have more resolution elements across the disk, to fully analyse the kinematics.} 
    \label{fig:FLS02}
\end{figure*}

\subsection{Kinematics}
\label{section:kinematics}

\begin{table}
	\centering
	\caption{Kinematic details from the best fit thin model of $^{\rm 3D}$BAROLO}
	\label{tab:kinematics}
	\begin{threeparttable}
	\begin{tabular}{ lcccc }
        \hline
        Object & PA\tnote{1} & Inclination\tnote{2}  &	V$_{\rm rot}$(outer ring)\tnote{3}	& $\sigma_{\rm gas}$(max)\tnote{4}  \\
         & degrees & degrees & km s$^{-1}$ & km s$^{-1}$ \\
        \hline\hline
        CDFS1-W	& 306	& 79	&  259 $\pm$ 14	& 42 $\pm$ 10	\\
        SWIRE5	& 188	& 50	& 346 $\pm$ 46 & 42 $\pm$ 6 \\
        \hline
    \end{tabular}
    \begin{tablenotes}
        \item[1] Position angle of the major axis on the receding side of the
        galaxy, taken anticlockwise from north. There are no errors as this value was not allowed to vary.
        \item[2] Inclination angle of the galaxy with respect to the observer, with 0$^{\circ}$ being face-on. There are no errors as this value was not allowed to vary.
        \item[3] The inclination corrected rotational velocity in the outer ring of the best fit $^{\rm 3D}$BAROLO model.
        \item[4] Maximum velocity dispersion in the best fit ring model of $^{\rm 3D}$BAROLO.
    \end{tablenotes}
    \end{threeparttable}
\end{table}

To fit a kinematic model to the CO(3-2) emission line we used the $^{\rm 3D}$BAROLO code, which fits a 3D tilted ring model to emission line data cubes (\citealp{di_teodoro_3d_2015}).
We used the SEARCH algorithm within $^{\rm 3D}$BAROLO to mask the galaxy from the background. The algorithm is a source finder that works by first making a 5$\sigma$ cut across all channels and then assumes that all 3$\sigma$ emission in adjoining channels/pixels are part of the same source (\citealp{whitling_duchamp_2012}). 

The model $^{\rm 3D}$BAROLO disk requires the following parameters: x and y positions, systematic velocity, rotational velocity (v$_{\rm rot}$), velocity dispersion ($\sigma_{\rm gas}$), radial velocity, position angle (PA), inclination, scale height of disk and gas surface density. The best fit model is degraded to the same resolution as the input cube by convolving it with a 2D Gaussian that has the same width and PA as the observed synthesized beam of the input cube.  

We chose the peak of the continuum as the centre point of each galaxy, set the radial velocity to zero, assumed a thin disk for the scale height, fixed the systematic velocity to zero at the centre of the emission and left the gas surface density to be estimated by the algorithm. We initially let the code make initial guesses for the PA and inclination, and then left them fixed at that value. The v$_{\rm rot}$ and $\sigma_{\rm gas}$ were left as free parameters for the code to fit. For our ring size we choose 0.3$^{\prime \prime}$ - 0.4$^{\prime \prime}$ to maximize the number of rings when fitting our source, while also ensuring an adequate number of spaxels within each ring to enable an adequate fit. All other settings within the algorithm were left as default (see \citealp{di_teodoro_3d_2015} for full details of the model settings).

To determine the best fit model we compared the output model's 0th, 1st and 2nd moment maps to the observed 0th, 1st and 2nd moment maps. If the PA and inclination greatly differed from the observed moment maps we manually fixed these values to match those of the observed maps, and re-ran the algorithm to find the v$_{\rm rot}$ and $\sigma_{\rm gas}$. In the four targets that have CO(3-2) emission line detections FLS02-N and FLS02-S show signatures of disturbed disks and were unable to be adequately fitted with a $^{\rm 3D}$BAROLO model; whereas CDFS1-W and SWIRE7 show ordered rotation and were fit with a model thin disk. The observed moment maps, best fit model moment maps, plus the residual maps are shown in Figures \ref{fig:CDFS1}, \ref{fig:SWIRE5} and \ref{fig:FLS02}, respectively, with the position velocity diagrams, along the major kinematic axis, of the fitted galaxies shown in Appendix \ref{appendix:pv}. The best fit parameters that $^{\rm 3D}$BAROLO was able to estimate/vary are given in Table \ref{tab:kinematics}. 
These will be discussed in more detail in Section \ref{section:triggers}.

\section{Discussion}
\label{section:discussion}

\subsection{Molecular Gas and SFR at 0.2 $<$ z $<$ 0.6}
\label{section:gasandsfr}


Studies have shown that the distance above the MS is linked to both an increased gas fraction and to the SFE of the gas (e.g. \citealp{scoville_evolution_2017}, \citealp{popesso_dust_2020}). To explore this further, it is instructive to compare our results with other U/LIRG samples, at similar redshifts, from the literature. \cite{combes_galaxy_2011} presented CO observations for a sample of ULIRGs at 0.2 $<$ z $<$ 0.6 which were previously detected at 60 $\mu$m from IRAS and ISO observations. These ULIRGs are undergoing starbursts and have higher dust temperatures than our sample (T$_{\rm dust}$ = 37 - 60K with an average of 48K), and we will refer to these as warm ULIRGs. \cite{lee_fine_2017} investigated a sample of LIRGs, at 0.25 $<$ z $<$ 0.65, that have lower luminosities than our sample (log(L$_{\rm IR}$ / L$_{\odot}$) = 11.2 - 11.6) and are located on, or just above, the upper boundary of the MS (hereafter referred to as cold LIRGs).

\begin{figure*}
  \centering
  \begin{subfigure}{0.5\textwidth}
     \includegraphics[width=1\linewidth]{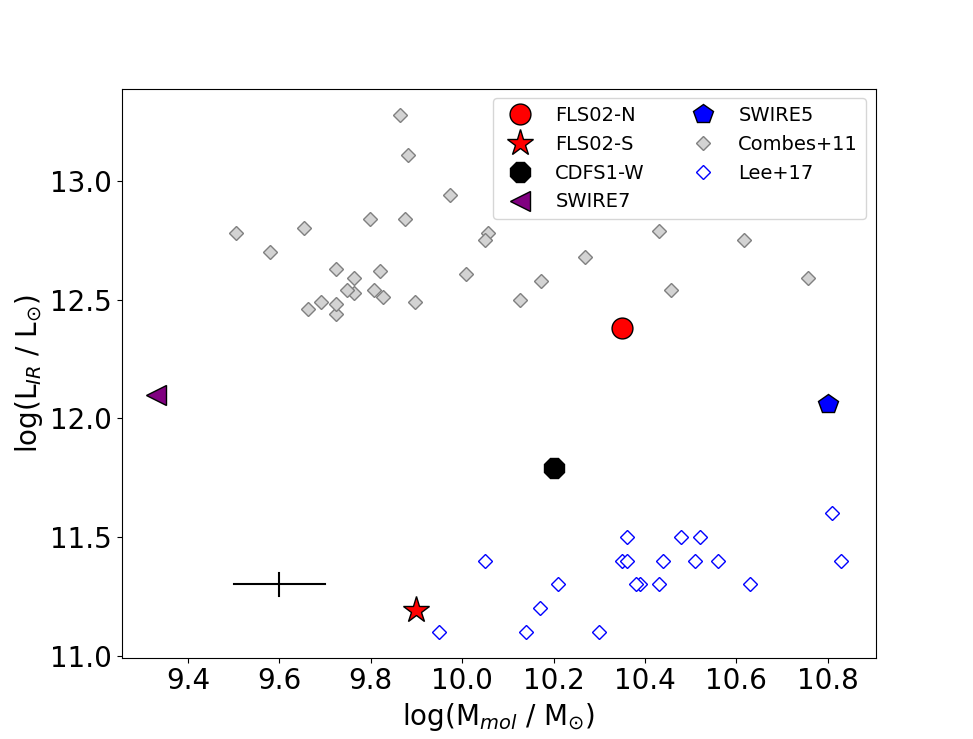}
   \end{subfigure}\hfil
   \begin{subfigure}{0.5\textwidth}
      \includegraphics[width=1\linewidth]{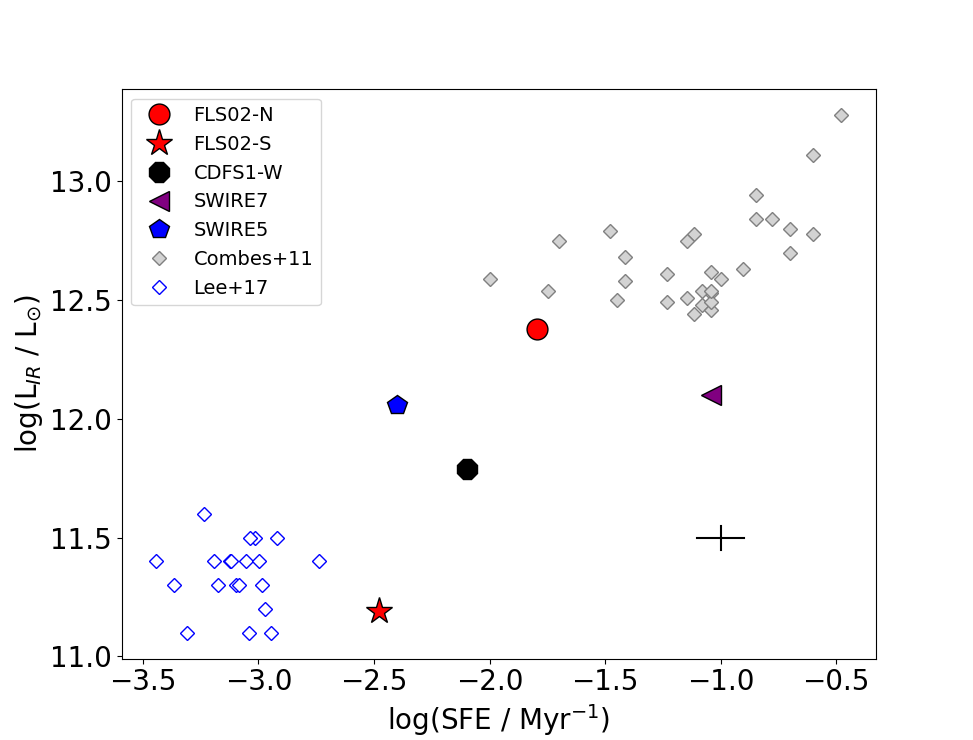}
   \end{subfigure}
   \caption{The total molecular gas mass, M$_{\rm mol}$ (left) and star formation efficiency, SFE (right) versus the IR luminosity, L$_{\rm IR}$, for our intermediate-z U/LIRGs. This is compared to cold LIRGs (blue diamonds, \citealp{lee_fine_2017}) and warm ULIRGs (grey diamonds, \citealp{combes_galaxy_2011}) from the same redshift. SWIRE7 was undetected in CO(3-2) and its M$_{\rm mol}$ and SFE are estimated from a 3$\sigma$ upper limit. The cross on both plots show the typical size of the error bars. These plots show that all three samples have similar M$_{\rm mol}$ reservoirs but they are forming stars at different rates of efficiency, indicating that higher SFR are being driven by increased SFE as opposed to a larger M$_{\rm mol}$ budget. }
   \label{fig:sfe}
\end{figure*}

Molecular gas masses and SFEs have been determined in both the cold LIRG and warm ULIRG samples (\citealp{combes_galaxy_2011}, \citealp{lee_fine_2017}), and these results along with our sample are plotted in Figure \ref{fig:sfe}. The SFE in \cite{combes_galaxy_2011} was presented in units of L$_{\odot}$ / M$_{\odot}$ so we converted it to SFR / M$_{\odot}$ by using the conversion factor of \cite{murphy_calibrating_2011}. For the $\alpha_{\rm CO}$ conversion factor \cite{combes_galaxy_2011} choose the value of 0.8 M$_{\odot}$ / (K km s$^{-1}$ pc$^{2}$) that is used in local ULIRGs and \cite{lee_fine_2017} applied the Milky Way value of 4.36 M$_{\odot}$ / (K km s$^{-1}$ pc$^{2}$). All three samples have similar M$_{\rm mol}$ so the amount of molecular gas available to form stars is not causing the differing L$_{\rm IR}$. When comparing SFE ($\tau_{\rm dep}$) of the three samples it is likely the increase in L$_{\rm IR}$ is being driven by the SFE of the molecular gas. There is a statistically significant strong positive correlation for log(SFE) - log(L$_{\rm IR}$) of the three samples, with a Pearson's $\rho$ = 0.91 (p-value < 0.001).

These results show that U/LIRGs located in the transition region between starbursts and MS galaxies in terms of dust temperature and L$_{\rm IR}$ also populate the transition region based on SFE, with an average $\tau_{\rm dep}$ of 150 Myr compared to $\tau_{\rm dep}$ $\approx$ 18 Myr for warm ULIRGs and $\approx$ 1 Gyr for cold LIRGs. If we expand the comparison to include the slightly higher redshift samples, \cite{freundlich_phibss2_2019} find an average $\tau_{\rm dep}$ of 0.84 Gyr for star forming galaxies at 0.5 $<$ z $<$ 0.8 (mean $\alpha_{\rm CO}$ = 4.0 M$_{\odot}$ / (K km s$^{-1}$ pc$^{2}$) based on using a metallicity scaling factor) and \cite{combes_gas_2013} measured an average $\tau_{\rm dep}$ $\approx$ 15 Myr for starbursts at 0.5 $<$ z $<$ 1 ($\alpha_{\rm CO}$ = 0.8 M$_{\odot}$ / (K km s$^{-1}$ pc$^{2}$)), both of which are consistent with the values at slightly lower redshifts.

Interestingly, \cite{lee_fine_2017} found a stronger correlation between distance above the MS and increased gas fraction, than between distance above the MS and SFE. With these findings in mind, it is possible, that an increase in just the gas fraction is sufficient to push a galaxy's SFR above the MS and into the transition region. However, a concurrent increase in SFE is required for a galaxy to approach the starburst regime, and we will investigate what may be triggering this heightened SFE in Section \ref{section:triggers}. 

Note that the values of SFE from various surveys and samples depends on the methodology of calculating the M$_{\rm mol}$. Hence, the SFE is implicitly determined by the choice of $\alpha_{\rm CO}$. As mentioned in Section \ref{section:gas}, local ULIRGs have an $\alpha_{\rm CO}$ up to five times smaller than the Milky Way. If we were to use the low-z ULIRG value ($\approx$ 0.8 M$_{\odot}$ / (K km s$^{-1}$ pc$^{2}$) for CO(1-0), \citealp{downes_rotating_1998}) our $\tau_{\rm dep}$ would be between 2 - 84 Myr (21 - 84 Myr if we remove SWIRE7). These $\tau_{\rm dep}$ values, from our four detections, are all greater than the average of 18 Myr (median of 12 Myr) determined for the warm ULIRGs. Due to the differing ISM conditions and merger stage in our sample compared to the local ULIRGs (M14, PS19) we stress that our choice of $\alpha_{\rm CO(3-2)}$(Z) is fully justified.

We reiterate that we were conservative in our choice of metallicity scaling factor. As discussed in Section \ref{section:gas}, we chose the scaling factor that gives us the smallest $\alpha_{\rm CO(3-2)}$(Z) and if we used the exact methodology of \cite{genzel_combined_2015} (taking the geometrical mean of two scaling factors) then we would get a larger M$_{\rm mol}$, with a corresponding slight increase in $\tau_{\rm dep}$, so the conclusions of this section are robust.

From Figure \ref{fig:sfe} it is clear the cold LIRGs and warm ULIRGs occupy two different nodes on the L$_{\rm IR}$ - SFE plane and appear to be bi-modal. Similarly, the local ULIRGS and MS galaxies occupy different regions of the $\Sigma_{\rm SFR}$ - $\Sigma_{\rm mol}$ plane, as seen in Figure \ref{fig:ks}. It is possible this bi-modality of the Kennicutt-Schmidtt relation may be due to selection effects as we move beyond local galaxies. MS galaxies make up an estimated 90\% of the star formation density between z = 0 - 2.5 (e.g. \citealp{tacconi_evolution_2020}), so most observations should fall on the locus of star forming galaxies on the Kennicutt-Schmidtt relation. When studying starbursts at higher-z we are biased towards the brightest targets which are easiest to detect and these bright objects are more likely to be associated with gas rich major mergers. The stochastic nature of mergers and limited time window to catch galaxies in an early interaction stage may mean that we currently do not have a large enough sample of galaxies that are transitioning between low and high SFE. With our sample of U/LIRGs, that straddles the transition region between regular star forming galaxies and starbursting galaxies, the SFE seems to show a more continuous change. \cite{shanngguan_interstellar_2019} studied local LIRGs in various stages of interaction and argued that they filled the gap between local MS galaxies and local starbursts suggesting a gradual change in SFE.
However, studies such as \cite{kennicutt_revisiting_2021} re-affirmed the bi-modality of the Kennicutt-Schmidtt relation based on local starburst and MS galaxies. As our sample is small we cannot draw any strong conclusions, and expanding the available samples of transitioning objects with future observations is necessary to fully explore this open issue.

\subsection{Triggers of Star Formation}
\label{section:triggers}

Both morphology and kinematics  provide important information on the mechanism that is triggering star formation within a galaxy. To investigate why our U/LIRGs are less efficient at forming stars than the starbursting ULIRGs of \cite{combes_galaxy_2011} but more efficient than the cold LIRGs of \cite{lee_fine_2017}, despite all having similar M$_{\rm mol}$, requires careful analysis of the kinematics within our targets. Local starbursts are associated with gas rich major mergers, with minor mergers and disk like morphology becoming more common with decreasing IR luminosity (e.g. \citealp{kartaltepe_multiwavelength_2010}, \citealp{bellocchi_distinguishing_2016}, \citealp{larson_morphology_2016}). Merger fraction has also been seen to increase with distance above the MS (e.g. \citealp{pearson_effect_2019}). In terms of the gas, it has been observed locally that both M$_{\rm mol}$ and SFE are increased during galaxy interactions (e.g. \citealp{pan_effect_2018}). Similar trends are seen in simulations (e.g. \citealp{moreno_spatially_2020}), although the boost in SFR due to interactions may not be large enough to push higher-z U/LIRGs above the MS (e.g. \citealp{martin_limited_2017}, \citealp{fensch_high-redshift_2017}). In terms of the availability of molecular gas to form stars, galaxies can have the same M$_{\rm mol}$ reservoirs but a different SFR due to stabilization by a compact and massive bulge preventing fragmentation into stars (e.g. \citealp{martig_morphological_2009}).
With these results in mind we try to determine how morphology and merger status is affecting SFR and SFE in our sample.

\subsubsection{CDFS1-W}
\label{section:cdfs1}

Despite CDSF1-W being separated by only $\approx$ 30 kpc from CDFS1-E, it has a clear rotating disk structure with little signs of disturbed rotation, so it must be at an early stage of interaction (see Figure \ref{fig:CDFS1}). This source has the lowest SFR$_{\rm IR}$ in our sample with 91 M$_{\odot}$ yr$^{-1}$, but does have the most compact star forming region with all obscured star formation located within the central $\approx$ 3 kpc of the disk. No obscured star formation is arising from the outer disk, therefore we suggest that the galaxy undergoes inside out star formation.
There is a signature of fast moving gas located in the central regions where star formation is taking place, as shown in the position velocity diagram in Figure \ref{fig:pv_CDFS1} with gas approaching 400 km s$^{-1}$. Outflows are ubiquitous in local U/LIRGs (e.g. \citealp{garcia-burillo_high-resolution_2015}, \citealp{pereira-santaella_spatially_2018}) and we need follow-up high spatial resolution observations of CDFS1-W to confirm if this fast moving gas is indeed an outflow.

\subsubsection{FLS02}
\label{section:fls02}
The FLS02 system has the largest SFR$_{\rm IR}$, at 380 M$_{\odot}$ yr$^{-1}$. The velocity fields (Figure \ref{fig:FLS02}) of both sources seems to show a disturbed velocity gradient, with an unresolved velocity dispersion profile. The pair of galaxies are separated by $\approx$ 20 kpc and the peak of the emission lines are offset by $\approx$ 150 km s$^{-1}$. This implies a more advanced merging stage than CDFS1. FLS02-N also has the shortest depletion time, $\tau_{\rm dep}$ = 62 Myr, in our sample, which is consistent with it being in a later interaction stage when compared to our other targets. Yet, both galaxies in FLS02 show signatures of being a rotating disk, which may explain why they are located in the transition region and do not yet have the SFE of a starburst. Higher spatial resolution observations are needed to properly probe the velocity structure and determine what is triggering the high rates of SFR.

\subsubsection{SWIRE5}
\label{section:swire5}
\begin{figure*}
  \centering
  \includegraphics[scale=0.5]{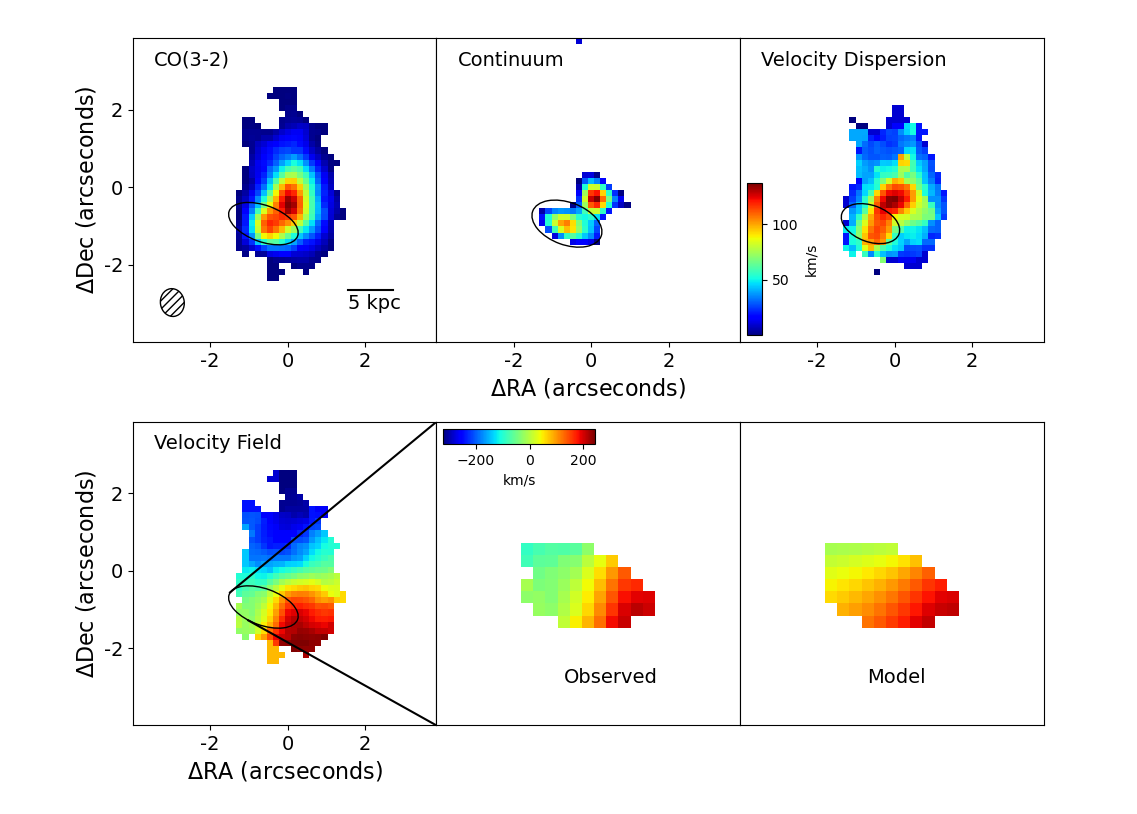}
  \caption{SWIRE5 shows excess flux, south east of the galaxy centre, than would be expected from an isolated disk as shown by the black ellipse in the top left plot. This corresponds to a second peak in the continuum, top middle, and a larger than expected gas velocity dispersion, rop right. There is also a slight deviation between the modelled and observed velocity field as shown in the bottom panels. Without higher spatial resolution data it is not clear what is causing this asymmetry. Possibilities include a hidden minor merger or a large disk instability.}
  \label{fig:merger}
\end{figure*}

SWIRE5's morphology and kinematics are an interesting case as it appears to be an isolated disk with a clear velocity gradient along a single axis (Figure \ref{fig:SWIRE5}). But, inspection of both the 0th, 1st and 2nd moment maps and the residual maps from the best fit disk model show an excess of CO(3-2) flux, south east of the galaxy centre, than would be expected from a rotating disk (see Figure \ref{fig:merger}) causing an irregular morphology. This excess flux is co-spatial to a secondary peak in the continuum, a region of high velocity dispersion and a small deviation between the modelled and observed velocity field.  
If we select a region similar in size to the second peak in the continuum, as plotted in the bottom row of Figure \ref{fig:merger}, there seems to be a velocity gradient along the major axis, when compared to the best fit model disk. 
PS19 also made moment maps and fit kinematics to SWIRE5 with the H$\alpha$ emission line (see Figure B6 in PS19). The 0th moment map of the H$\alpha$ line does not show a similar asymmetry in the flux map but the velocity dispersion shows similar high values in the south-east region of the galaxy. These observations were taken by the SWIFT telescope and due to the seeing limitations (the seeing FWHM for SWIRE5 was 2.7$^{\prime\prime}$), it does not provide higher spatial resolution data compared to our NOEMA observations.

This disturbance may be an unresolved region of clumpy star formation caused by disk instabilities (e.g. \citealp{behrendt_possible_2019}) or another possibility is that we are seeing interacting galaxies hiding behind each other and appearing as one, such as that potentially observed in optical images of a low-z system in \cite{ciraulo_two_2021} . The hidden merger of \cite{ciraulo_two_2021} was taken from a sample of galaxies with double peak emission line profiles (\citealp{maschmann_ultimate_2019}, \citealp{maschmann_double-peak_2020}) some of which have been suggested to harbour hidden minor mergers. SWIRE5 has an asymmetric double peak line profile, as shown in Figure \ref{appendix:spectra}, which is additional circumstantial evidence for a hidden minor merger. Conversely, the $\tau_{\rm dep}$ of 251 Myr suggest that if there is a hidden minor merger it has not boosted the SFE of the gas, compared to the known interacting galaxies in our sample, although the $\tau_{\rm dep}$ is much higher than isolated disks, such as the LIRGs of \cite{lee_fine_2017} with typical values of 1 Gyr/ In either eventuality, this disturbance is co-spatial with a region of ongoing obscured star formation, so whatever underlying mechanism is causing the asymmetry in the CO(3-2) emission is also triggering star formation.

As the major axis of the synthesized beam is $\approx$ 3.5 kpc in width we require high spatial resolution observations to disentangle the possible scenarios. Signatures of minor mergers will also be explored using mock observations from simulations in Hogan et al. (in prep), which will help interpret this and similar types of observations.
\bigskip

To conclude, all four galaxies that are observed in CO(3-2) are interacting and/or show deviations from a model isolated disk galaxy, which may explain why their SFE is greater than cold LIRGs at similar redshifts. None of the four appear to be major mergers, so it is likely that the gradual rise in SFE, from the least luminous LIRGs to the highest luminosity ULIRGs at intermediate redshifts, is associated with the degree of isolation versus interaction.

\subsection{CDFS1-E \& SWIRE7 Non-Detections}
\label{section:swire7}

An important consideration when measuring the molecular gas content of galaxies is that direct observation of cold H$_{\rm 2}$ gas is difficult as the lowest energy transitions are rotational quadrupole transitions which require gas temperatures $>$ 100K, and this is higher than a typical temperature in the molecular ISM. The next most abundant molecule is CO, whose low rotational excitation energy and critical density have made it the traditional M$_{\rm mol}$ tracer (e.g. \citealp{carilli_cool_2013}, \citealp{combes_molecular_2018}). However, CO can fail to track the total H$_{\rm 2}$ reservoir in certain environments, such as the metal poor ISM of dwarf galaxies, where the requisite metals are not available to shield the CO gas from ultra-violet (UV) photons leading to photo-dissociation (e.g. \citealp{papadopoulos_molecular_2002}, \citealp{wolfire_dark_2010}). H$_{2}$ has a greater capacity for self-shielding from UV photons so a non-negligible amount of molecular gas may be available to form stars, despite being deficient in CO, and has been dubbed CO-dark gas (e.g. \citealp{krumholz_which_2011}).

Although CDFS1-E is observed in H$\alpha$ in PS19 it is not detected in either CO(3-2) or continuum. This galaxy has the lowest metallicity amongst all the galaxies in PS19, so it is possible that CDFS1-E is a CO-dark dwarf galaxy, and that the large L$_{\rm IR}$ is arising entirely from CDFS1-W. Without further observations we cannot offer a definitive conclusion for the nature of this galaxy.

\begin{figure*}
  \centering
  \includegraphics[scale=0.4]{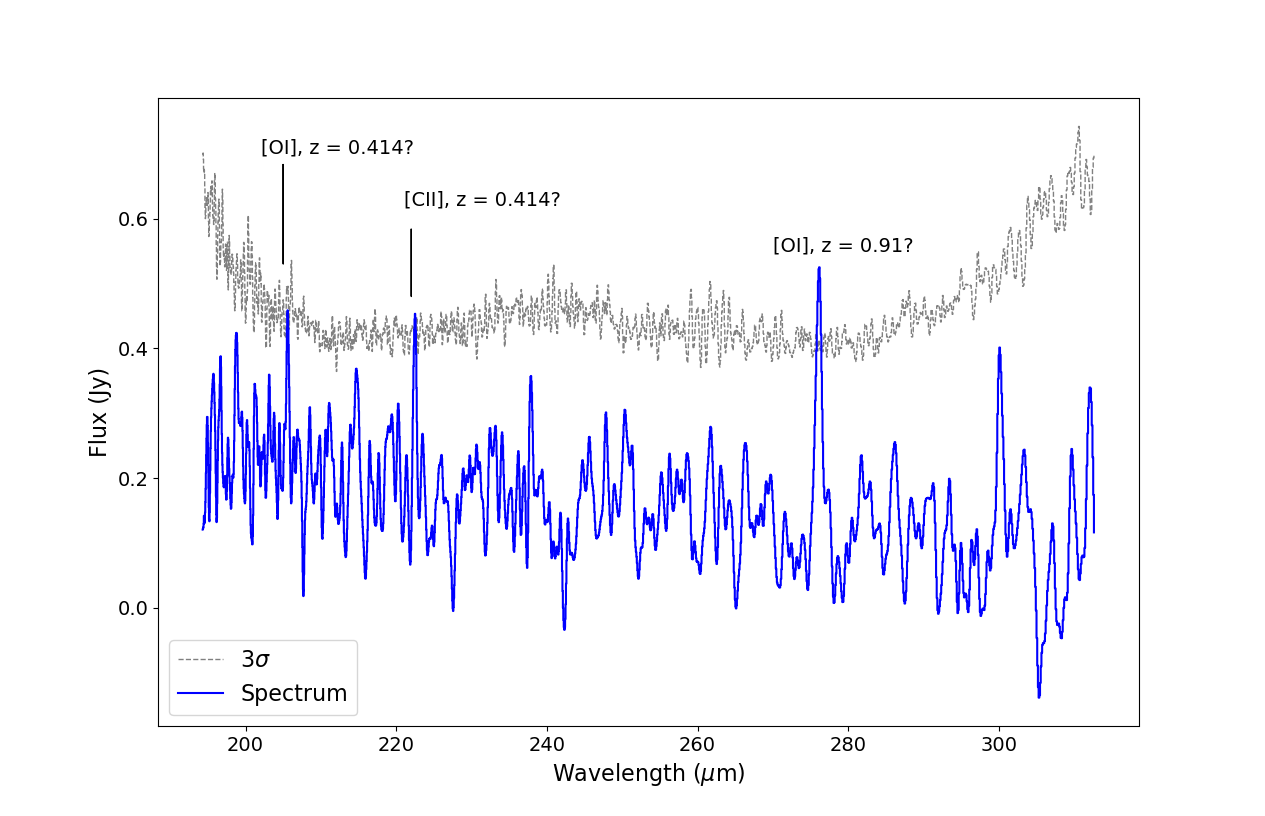}
  \caption{The spectrum of SWIRE7 taken by SPIRE-FTS. This galaxy was undetected in CO(3-2) but has a robust detection of an emission line at 927.8 nm in PS19 which is assumed to be the H$\alpha$ line at z = 0.414 (see Section \ref{section:swire7}  for full details). There are marginal 3$\sigma$ detections at 222 and 205 $\mu$m, which agree with both [CII] 158 $\mu$m and [OI] 145 $\mu$m lines at z = 0.414, corroborating PS19's H$\alpha$ detection. But the possible line at  277$\mu$m fits with [OI] 145 $\mu$m at z = 0.91 and PS19's detection being H$\beta$, but it is unlikely a strong H$\beta$ detection would be observed in a ULIRG at z = 0.91.}
  
  \label{fig:spire-fts}
\end{figure*}

SWIRE7 poses more of a challenge as it has a strong ($\approx$ 20$\sigma$) detection in the continuum despite having no CO(3-2) detection. PS19 has a robust detection of an emission line at 927.8 nm with the most likely identification to be H$\alpha$ at z = 0.414. Having re-examined the SPIRE-FTS spectrum presented in M14 there are possible 3$\sigma$ detections of both [CII] 158 $\mu$m and [OI] 145 $\mu$m lines at the expected wavelength, observed at 222 and 205 $\mu$m respectively, as shown in Figure \ref{fig:spire-fts}. 
The only other possibility is that PS19's observed emission line is H$\beta$ at z = 0.91, with the $\approx$ 3$\sigma$ peak at 277$\mu$m being the [OI] 145 $\mu$m line. This does not fit with the 222 and 205 $\mu$m lines and a strong detection of H$\beta$ in a z = 0.91 ULIRG seems unlikely.

It may be that this galaxy has a metal poor ISM, corroborated by the lack of [NII] 6583 \AA\ detection in PS19. However, the SFR$_{\rm IR}$ of 190 M$_{\odot}$ yr$^{-1}$ does not agree with a classification as a dwarf galaxy.  
Alternatively, the continuum detection may be due in part to thermal free-free emission from electrons scattering in the HII regions (e.g. \citealp{condon_essential_2016}), as opposed to being the tail of the Rayleigh-Jeans continuum, which would partly explain the 20$\sigma$ continuum detection despite being a metal poor object.

Recently, possible examples (or cases) of CO-dark gas have been observed in a small number of intermediate-z LIRGs which have been detected in [CI] and dust continuum but are deficient in CO (\citealp{dunne_dust_2021}). Although we do not have a [CI] detection, if the marginal detection of [CII] is correct then the galaxy may not be carbon deficient but may have some mechanism in the ISM that is rapidly dissociating CO into [CI] and [CII] (e.g. \citealp{bisbas_cosmic-ray_2017}).

If SWIRE7 is a true CO-dark ULIRG, and not an incorrect redshift, then it poses a challenge to the understanding of the ISM within dusty star forming galaxies galaxies, so followup observations are required.

\section{Conclusion}
\label{section:conclusion}
We have carried out CO(3-2) observations with NOEMA and ALMA of six targets (four systems) located at 0.28 $<$ z $<$ 0.44. Our sample consists of U/LIRGs that straddle the transition region between regular star forming galaxies and starbursts. Our sample also fills the gap between previous studies of low luminosity LIRGs and high luminosity ULIRGs at this epoch, and allows us to explore the entire population of luminous IR galaxies at a time when the universe is experiencing a rapid decline in star formation rate density. The main results of this paper are:
\begin{enumerate}
\item We have detected CO(3-2) emission in four of our targets: CDFS1-W, FLS02-N\&S and SWIRE5. There were two CO(3-2) non-detections: CDFS1-E and SWIRE7, but despite this a strong continuum detection was found in SWIRE7. These observations show our sample of intermediate-z U/LIRGs fall between the loci of MS galaxies and starbursts on the Kennicutt-Schmidtt relation, suggesting that galaxies in the transition region between starbursts and MS galaxies also populate the transition region on the $\Sigma_{\rm CO}$ - $\Sigma_{\rm IR}$ plane.
\item Our sample of galaxies have a $\tau_{\rm dep}$ on the order of 100 Myr and comparing to U/LIRGs at a similar redshift, the depletion time is lower than cold LIRGs, with a $\tau_{\rm dep}$ on the order of 1 Gyr, but higher than warm starbursting ULIRGs, with $\tau_{\rm dep}$ of approximately 10 Myr. All three U/LIRG samples have similar M$_{\rm mol}$ budgets but differing star formation efficiencies. A strong positive correlation between log($L_{\rm IR}$) and log(SFE) suggests it is an increasing SFE that is driving the increasing SFRs, as opposed to larger gas fractions. The location of our sample on the log($L_{\rm IR}$) and log(SFE) plane points to a gradual change in SFE as a galaxy transitions between MS and starburst. 
\item All four galaxies for which we detected a CO(3-2) emission line show signatures of rotating disks. Two of these show ordered rotation (CDFS1-W and SWIRE5) that could be adequately fit with a model rotating gas disk using $^{\rm 3D}$Barolo. This is indicative of our galaxies being isolated or in an early stage of interaction.
\item As both FLS02 and CDFS1 are interacting systems and SWIRE5 shows a disturbance that deviates from the disk model, which may be a hidden minor merger or a disk instability, it seems that morphology and kinematics play a significant role in elevating the SFE of a galaxy above the MS.
\end{enumerate}
In summary, our sample of transitioning U/LIRGs fill the gap between main sequence galaxies and starbursts on the Kennicutt-Schmidtt relation. Their SFE bridges the gap between regular star forming galaxies and starbursts at intermediate-z and suggest a continuous change in SFE between these two populations. These differences in SFE may arise due to morphology and interaction stage.

\section*{Acknowledgements}
We would like to thank the referee for their insightful feedback and helping us improve the quality of this paper. We are thankful to Michael Bremer and Orsolya Feher for helping reduce the NOEMA data.
DR and IGB acknowledge support from STFC through grant ST/S000488/1. 
NT acknowledges support from STFC through grants ST/N002717/1 and ST/S001409/1.
MPS acknowledges support from the Comunidad de Madrid through Atracci\'on de Talento Investigador Grant 2018-T1/TIC-11035. 
SGB and AAH acknowledge support from research grant PGC2018-094671-B-I00 (MCIU/AEI/FEDER,UE).
SGB acknowledges support from the research project PID2019-106027GA-C44 of the Spanish Ministerio de Ciencia e Innovaci\'on.
This paper makes use of the following ALMA data: ADS$/$JAO.ALMA\#2016.1.00896.S. ALMA is a partnership of ESO (representing its member states), NSF (USA) and NINS (Japan), together with NRC (Canada), MOST and ASIAA (Taiwan), and KASI (Republic of Korea), in cooperation with the Republic of Chile. The Joint ALMA Observatory is operated by ESO, AUI$/$NRAO and NAOJ. This work is based on observations carried out under project number W19BV and W20GC with the IRAM NOEMA Interferometer. IRAM is supported by INSU$/$CNRS (France), MPG (Germany) and IGN (Spain).
DR and IGB acknowledge support from STFC through grant ST/S000488/1. 
NT acknowledges support from STFC through grants ST/N002717/1 and ST/S001409/1.
MPS acknowledges support from the Comunidad de Madrid through Atracci\'on de Talento Investigador Grant 2018-T1/TIC-11035. 
SGB and AAH acknowledge support from research grant PGC2018-094671-B-I00 (MCIU/AEI/FEDER,UE).

\section*{Data Availability}
 
The ALMA data underlying this research is available from the ALMA archive. The NOEMA data can be shared upon reasonable request to the corresponding author.  




\bibliographystyle{mnras}
\bibliography{library} 



\appendix
\onecolumn

\section{Optical Images of Our Sample.}
Taken from \cite{pereira-santaella_optical_2019}. Note that north points down in the CDFS1 and FLS02 images and left the SWIRE5 images (north in all CO images points up) as seen by the arrows in the bottom left corner.
\label{appendix:opt}

\begin{figure}[H]
  \centering
  \includegraphics[scale=0.65]{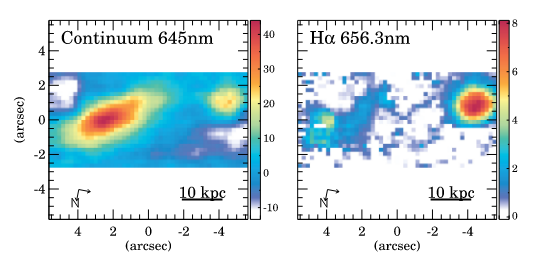}
  \caption{H$\alpha$ and rest frame 645 nm continuum for CDFS1, taken by the SWIFT telescope. The units for the continuum and emission line maps are 10$^{-15}$ erg s$^{-1}$ cm$^{-2}$ $\mu$m$^{-1}$and 10$^{-17}$ erg s$^{-1}$ cm$^{-2}$ respectively. The two arrows in the lower left-hand corner show the North and East directions. The seeing FWHM for this source was 1.4$^{\prime\prime}$/ 6 kpc. Note that north points down, opposite to our CO image of CDFS1.} 
  \label{fig:CDFS1_optical}
\end{figure} 

\begin{figure}[H]
  \centering
  \includegraphics[scale=0.65]{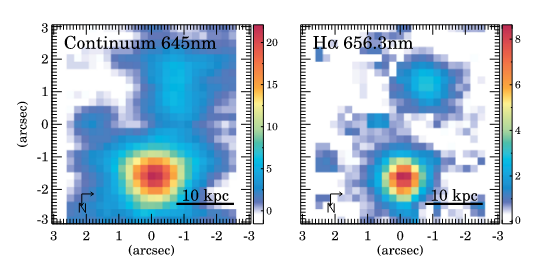}
  \caption{Same as Image \ref{fig:CDFS1_optical} but for FLS02. The seeing FWHM for this source was 1.6$^{\prime\prime}$/ 9 kpc. Note that north points down, opposite to our CO image of FLS02.} 
  \label{fig:FLS02_optical}
\end{figure} 

\begin{figure}[H]
  \centering
  \includegraphics[scale=0.60]{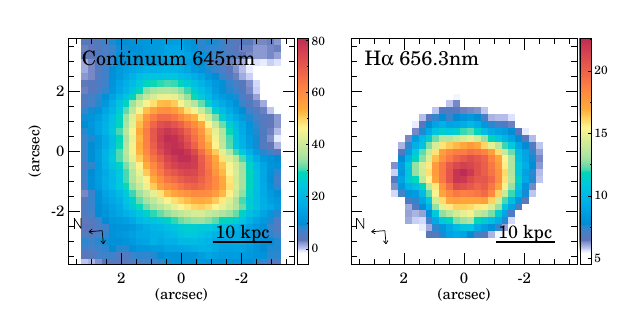}
  \caption{Same as Image \ref{fig:CDFS1_optical} but for SWIRE5, The seeing FWHM for this source was 2.7$^{\prime\prime}$/ 13.7 kpc. Note that north points left.} 
  \label{fig:SWIRE5_optical}
\end{figure} 

\begin{figure}[H]
  \centering
  \includegraphics[scale=0.75]{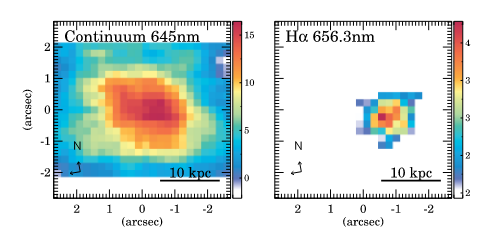}
  \caption{Same as Image \ref{fig:CDFS1_optical} but for SWIRE7. The seeing FWHM for this source was 2.1$^{\prime\prime}$/ 11 kpc.} 
  \label{fig:SWIRE7_optical}
\end{figure}

\section{The Integrated Spectra for our Four CO(3-2) Detections.}
\label{appendix:spectra}
\vspace{-0.5cm}

\begin{figure}[H]
  \centering
  \begin{subfigure}{0.45\textwidth}
     \includegraphics[width=\linewidth]{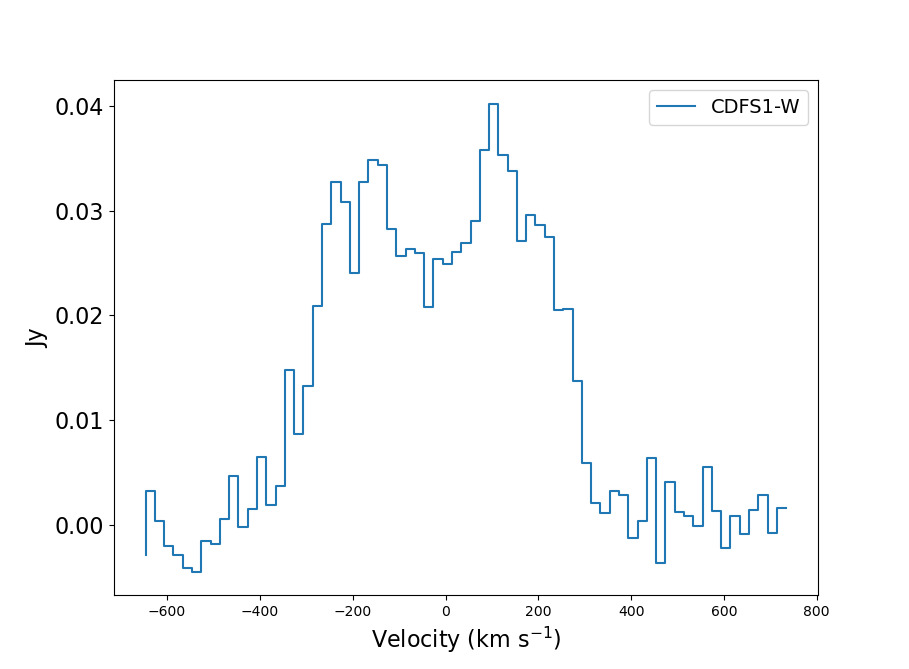}
   \end{subfigure}\hfil
   \begin{subfigure}{0.45\textwidth}
      \includegraphics[width=\linewidth]{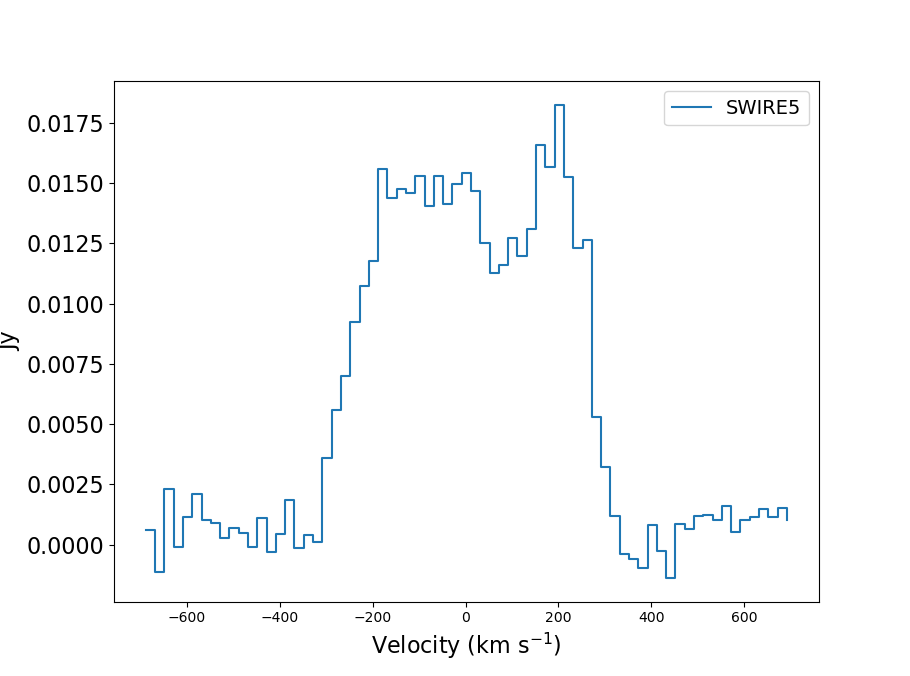}
   \end{subfigure}
   \begin{subfigure}{0.45\textwidth}
     \includegraphics[width=\linewidth]{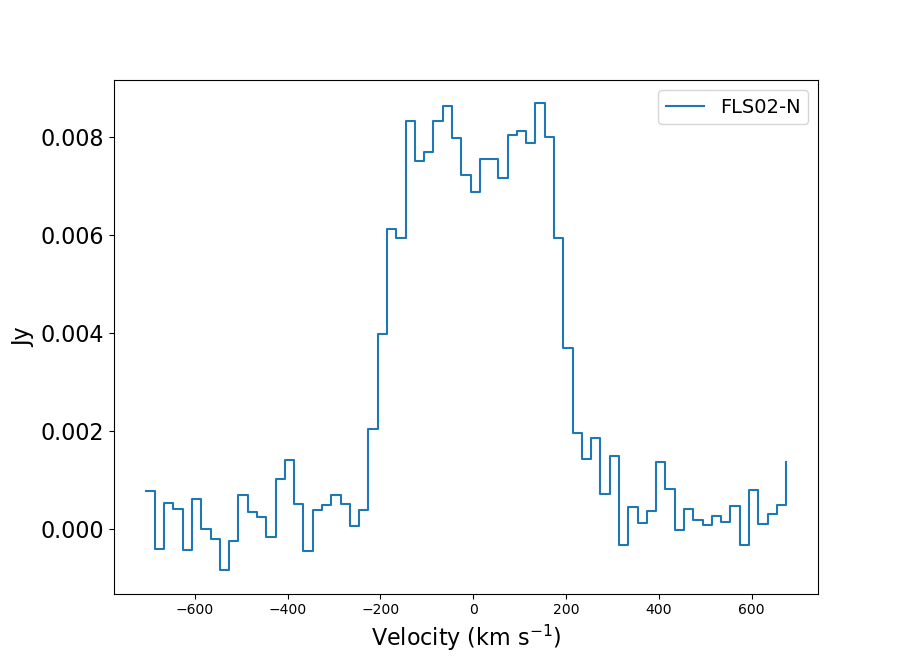}
   \end{subfigure}\hfil
   \begin{subfigure}{0.45\textwidth}
      \includegraphics[width=\linewidth]{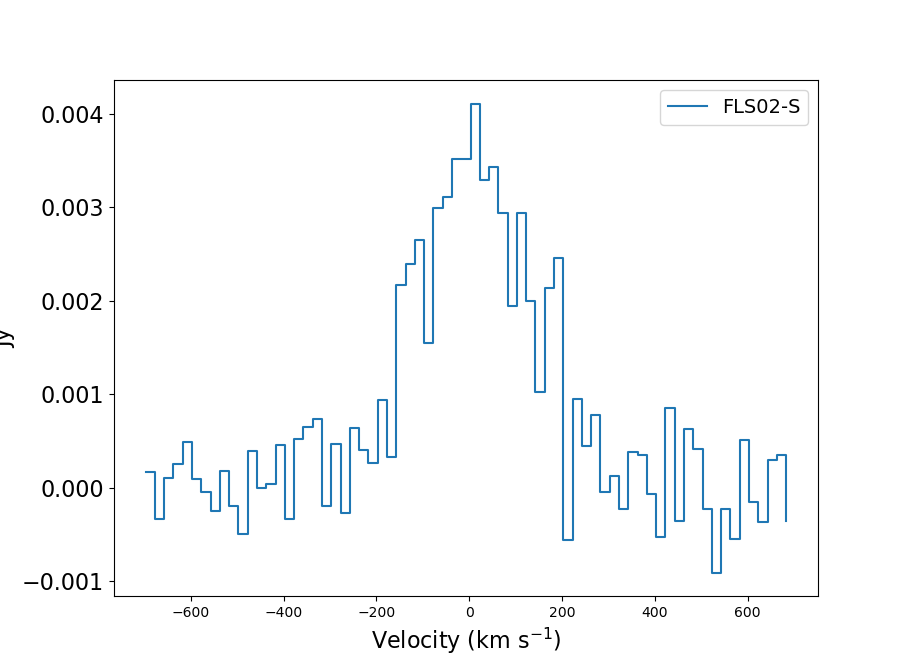}
   \end{subfigure}
   \caption{The galaxy integrated spectra of each of our targets with a CO(3-2) detection: CDFS1-W (top left), SWIRE5 (top right), FLS02-N (bottom left) and FLS02-S (bottom right).} 
  \label{fig:spectra}
\end{figure}

\section{Expected Position of CDFS1-E with respect to CDFS1-W}
\label{appendix:companion}

\begin{figure}[H]
  \centering
  \includegraphics[scale=0.45]{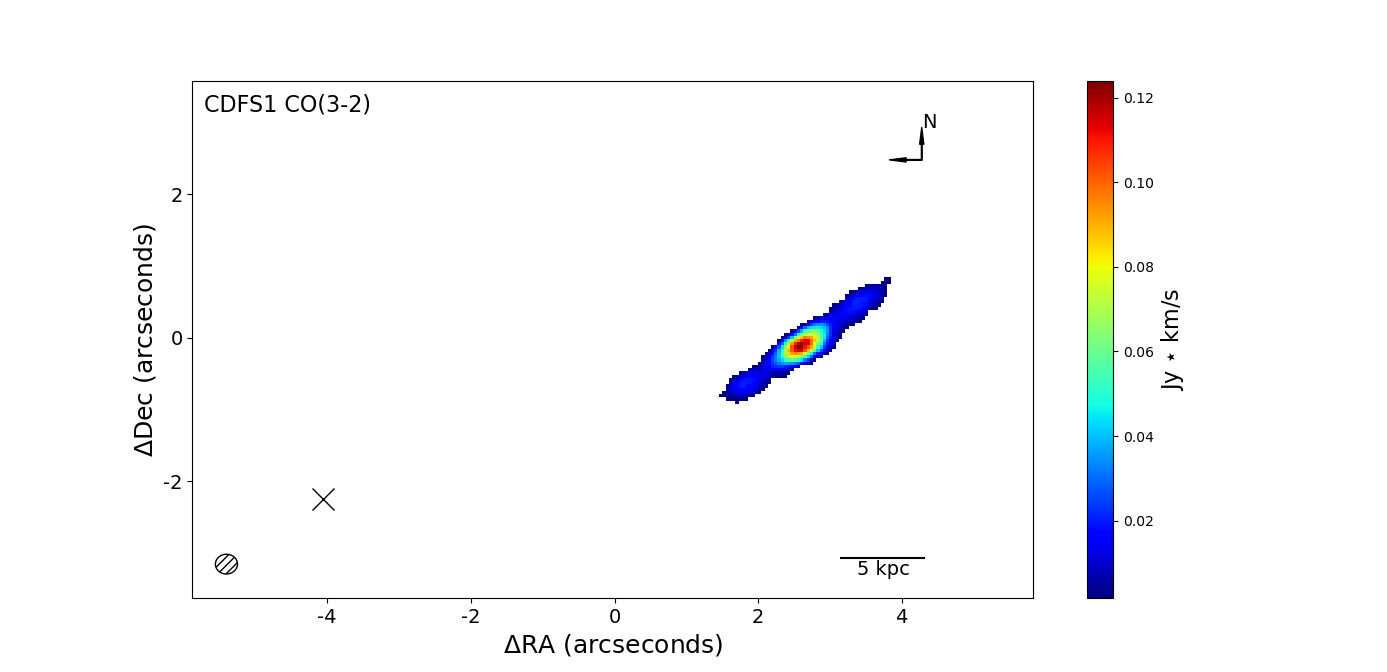}
  \caption{CO(3-2) intensity map of the CDFS1 system, in units if Jy km s$^{-1}$, with the expected position of the undetected CDFS1-E marked with an X. The beam size is shown in the bottom left corner with the physical scale shown in the bottom right.} 
  \label{fig:companion}
\end{figure} 

\section{Position Velocity Diagrams along the Major Kinematic Axis for our Two Targets fitted with a model Disk}
\label{appendix:pv}

\begin{figure}[H]
  \centering
  \includegraphics[scale=0.3]{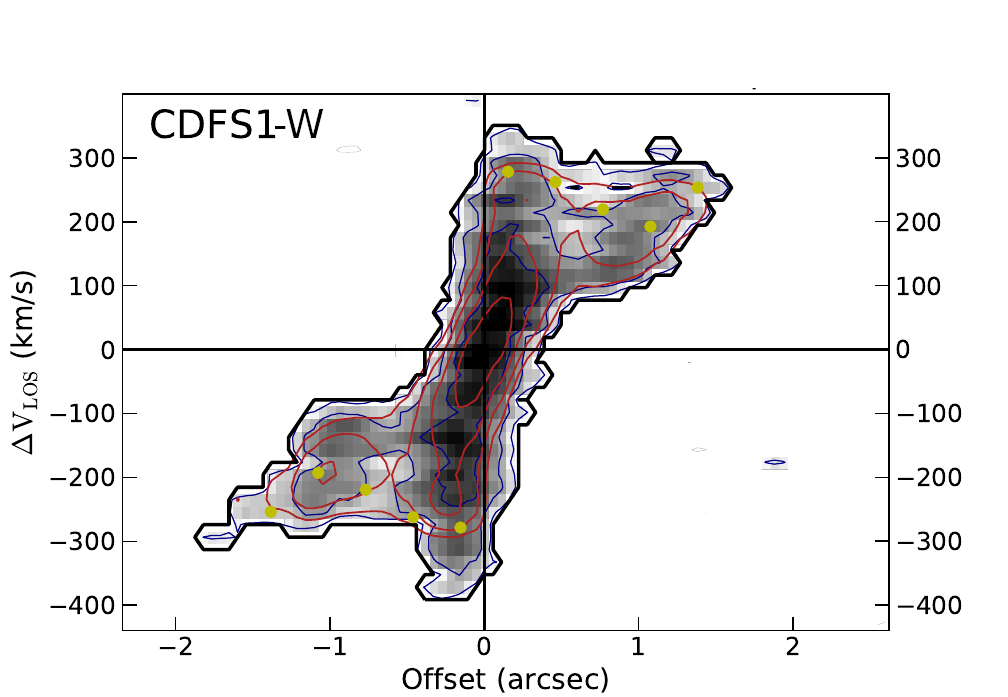}
  \caption{Position velocity diagram along the major kinematic axis of CDFS1-W. The black contours show the observed galaxy with the red overlaid contours represent the model galaxy from $^{\rm 3D}$Barolo. The yellow points are the V$_{\rm rot}$ calculated for each ring in the model disk. There seems to fast moving gas within the inner 0.5$^{\prime \prime}$ of the disk with projected velocities up to $\approx$ 400 km s$^{-1}$, which may be suggestive of an outflow.} 
  \label{fig:pv_CDFS1}
\end{figure} 

\begin{figure}[H]
  \centering
  \includegraphics[scale=0.3]{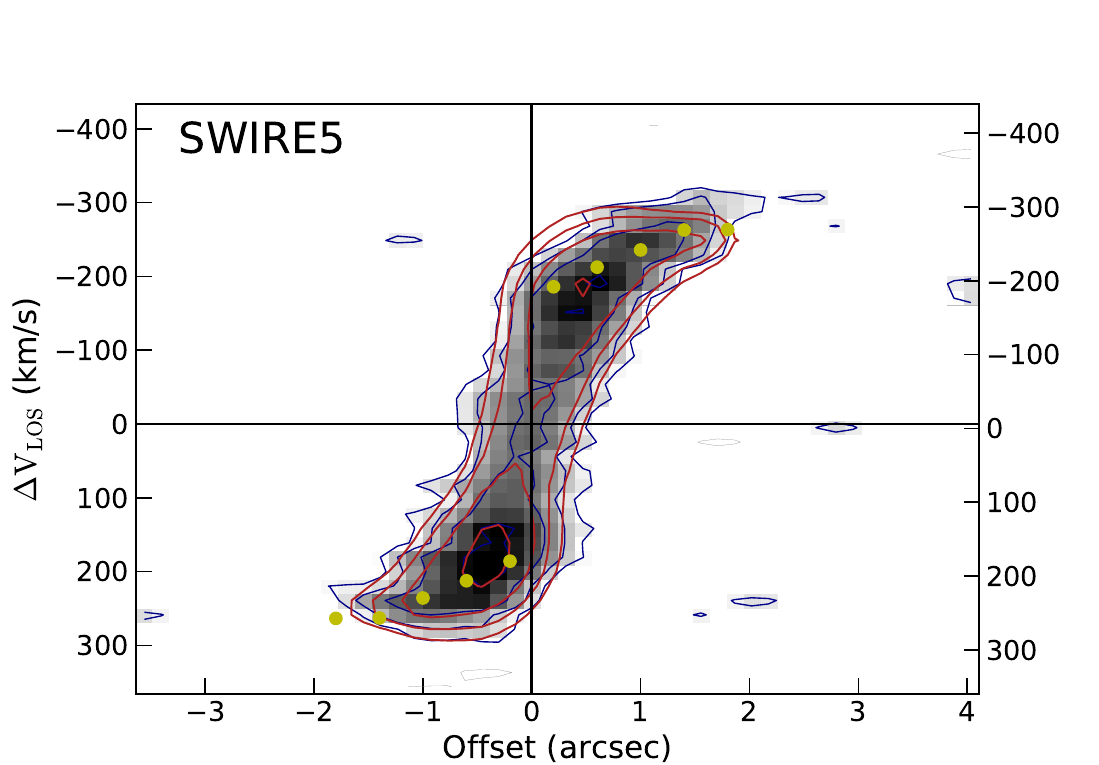}
  \caption{Position velocity diagram along the major kinematic axis of SWIRE5, with the black contours show the observed galaxy with the red overlaid contours represent the model galaxy from $^{\rm 3D}$Barolo. The yellow points are the V$_{\rm rot}$ calculated for each ring in the model disk..} 
  \label{fig:pv_SWIRE5}
\end{figure}


\bsp	
\label{lastpage}
\end{document}